\documentclass[journal]{IEEEtran}

\usepackage[switch]{lineno}
\usepackage{amsmath,amsfonts,graphicx,float}
\usepackage{array}
\usepackage{cleveref}
\usepackage{booktabs}
\usepackage{ragged2e}
\usepackage{multicol}
\usepackage{multirow,bigstrut}
\usepackage{color}
\usepackage{xspace}
\usepackage{graphicx}
\usepackage[justification=centering]{caption}
\usepackage{subcaption}
\usepackage{lipsum}
\usepackage{cite}
\usepackage{blindtext}
\usepackage{multicol}
\usepackage{hhline}
\usepackage{bm}
\usepackage{epstopdf}

\newcolumntype{L}{>{\centering\arraybackslash}m{0.035\textwidth}}
\newcolumntype{Q}{>{\centering\arraybackslash}m{0.048\textwidth}}
\newcolumntype{M}{>{\centering\arraybackslash}m{0.100\textwidth}}
\newcolumntype{N}{>{\centering\arraybackslash}m{0.050\textwidth}}
\newcolumntype{S}{>{\centering\arraybackslash}m{0.075\textwidth}}
\newcolumntype{K}{>{\centering\arraybackslash}m{0.15\textwidth}}
\newcolumntype{O}{>{\centering\arraybackslash}m{0.065\textwidth}}
\newcolumntype{J}{>{\centering\arraybackslash}m{0.2\textwidth}}
\newcolumntype{F}{>{\centering\arraybackslash}m{0.045\textwidth}}

\begin{document}

\title{Streaming Video QoE Modeling and Prediction: A Long Short-Term Memory Approach}

\author{
Nagabhushan~Eswara$^1$,~\textit{Student~Member},~\textit{IEEE},~Ashique~S$^2$,~Anand~Panchbhai$^3$,~Soumen~Chakraborty$^4$, \\
~Hemanth~P.~Sethuram$^4$,~Kiran~Kuchi$^2$,~Abhinav~Kumar$^2$,~\textit{Member},~\textit{IEEE},\\~and~Sumohana~S.~Channappayya$^1$,~\textit{Member},~\textit{IEEE}

\thanks{
$^1$N.~Eswara and S.~S.~Channappayya are with the Laboratory for Video and Image Analysis (LFOVIA), Department of Electrical Engineering, Indian Institute of Technology Hyderabad, Kandi, Telangana 502285, India.
(email: ee13p1004@iith.ac.in; sumohana@iith.ac.in)
}

\thanks{
$^2$A.~S,~K.~Kuchi, and A.~Kumar are with the Department of Electrical Engineering, Indian Institute of Technology Hyderabad, Kandi, Telangana 502285, India.
(email: ee14btech11003@iith.ac.in; kkuchi@iith.ac.in; abhinavkumar@iith.ac.in)
}

\thanks{
$^3$A.~Panchbhai is with the Department of Computer Science and Engineering, Indian Institute of Technology Bhilai, Sejbahar, Raipur, Chhattisgarh 492015, India.
(email: anandp@iitbhilai.ac.in)
}

\thanks{
$^4$S.~Chakraborty and H.~P.~Sethuram are with Intel Corporation, Bengaluru, Karnataka 560103, India.
(email: soumen.chakraborty@intel.com; hemanth.p.sethuram@intel.com)
}

}

\date{\today}
\maketitle

\begin{abstract}
HTTP based adaptive video streaming has become a popular choice of streaming due to the reliable transmission and the flexibility offered to adapt to varying network conditions.
However, due to rate adaptation in adaptive streaming, the quality of the videos at the client keeps varying with time depending on the end-to-end network conditions. 
Further, varying network conditions can lead to the video client running out of playback content resulting in rebuffering events.
These factors affect the user satisfaction and cause degradation of the user quality of experience (QoE). 
It is important to quantify the perceptual QoE of the streaming video users and monitor the same in a continuous manner so that the QoE degradation can be minimized. 
However, the continuous evaluation of QoE is challenging as it is determined by complex dynamic interactions among the QoE influencing factors.
Towards this end, we present LSTM-QoE, a recurrent neural network based QoE prediction model using a Long Short-Term Memory (LSTM) network. The LSTM-QoE is a network of cascaded LSTM blocks to capture the nonlinearities and the complex temporal dependencies involved in the time varying QoE.
Based on an evaluation over several publicly available continuous QoE databases, we demonstrate that the LSTM-QoE has the capability to model the QoE dynamics effectively. 
We compare the proposed model  with the state-of-the-art QoE prediction models and show that it provides superior performance across these databases.
Further, we discuss the state space perspective for the LSTM-QoE and show the efficacy of the state space modeling approaches for QoE prediction.

\end{abstract}


\section{Introduction}
\label{sec:intro}

Streaming videos on demand over Hyper Text Transfer Protocol (HTTP) has grown significantly in the recent years. According to Cisco's Visual Networking Index \cite{cisco}, mobile video traffic accounted for 60\% of the total mobile data traffic in 2016. It is estimated that videos will constitute more than three-fourths of the world's mobile data traffic by the year 2021. Such a massive growth in the video traffic will lead to a tremendous amount of stress on the video delivery infrastructure. Therefore, it is important for the network service providers to perform a careful and optimal utilization of the available resources for video streaming while maintaining an acceptable level of Quality-of-Experience (QoE) for the video users. 

Adaptive streaming solutions such as Dynamic Adaptive Streaming over HTTP, popularly known as DASH, provide an operative framework for media streaming over networks \cite{sodagar2011mpeg}. DASH has become a popular choice of media streaming as most networks today are configured to operate over Transfer Control Protocol (TCP) in HTTP/TCP. 
The media delivery in DASH is lossless as TCP is a reliable transport level protocol. However, network impairments such as congestion, poor wireless channel etc. can cause packet loss in the network resulting in significant delays in the packet arrival.
Delays in the video packet arrival can result in the emptying of the playback buffer causing the playback to stall. Such events are referred to as rebuffering events \cite{Vinay_Joseph}. The playback is not resumed until sufficient content is available in the buffer. 
The rebuffering events can also occur in wireless networks where the resources are limited and are shared between multiple users. Due to resource sharing, some users can end up being starved of resources, thereby reducing their throughput. The data rate of a wireless video user is highly influenced by network dynamics such as the number of users, load on the network and so on.
In order to minimize the occurrence of rebuffering events for video users, DASH provides an adaptive streaming capability to the clients (video users) to adapt their video rate in accordance with the changing network conditions. Rate adaptation is a key feature of adaptive streaming that is useful in dynamic and varying transmission environments such as mobile wireless networks. However, it should be noted that the videos encoded at different rates offer different video qualities. Hence, rate adaptation can result in a video quality that keeps varying with time. 
The QoE as perceived by the user is determined by a complex interplay of the time varying video quality and rebuffering events \cite{LFOVIA_QoE}. Given these dynamically varying QoE influencing factors in a video streaming session, the QoE evolution is continuous, dynamic, and time varying in nature.

Continuous time QoE monitoring is vital for optimizing the utilization of shared resources and thereby maximize the QoE of video users in the network.
It is also useful 
for performing optimal video rate adaptation
at the client so that the degradations in the QoE caused due to time varying quality as well as rebuffering events are minimized. 
While continuous QoE evaluation approaches such as \cite{LFOVIA_QoE}, \cite{NARX}, and \cite{Deepti_contQoE} have shown to provide a reasonable QoE prediction performance upon the databases over which they are trained, their prediction performance typically degrades when evaluated on other databases. For instance, the nonlinear autoregressive (NARX) model proposed in \cite{NARX} is shown to perform well over the LIVE Netflix Database \cite{LIVE_Netflix}, but yields a lower performance when evaluated over the LFOVIA QoE Database \cite{LFOVIA_QoE}. There is a need for a comprehensive QoE prediction model that performs consistently well across the QoE databases.
Furthermore, there is a need for improvement in the QoE prediction as compared to the existing QoE evaluation methods.
These form the motivating factors for this work. 

Therefore, in this paper, we present LSTM-QoE, a novel method for predicting the continuous QoE of video streaming users based on Long Short Term-Memory (LSTM).
We rely on LSTMs for QoE evaluation, as LSTMs have shown to be effective in modeling complex temporal dependencies in applications such as sequence labeling \cite{sequence_labeling}, visual recognition \cite{LSTM_recognition} etc.
The proposed LSTM-QoE model relies on three input features for continuous QoE prediction, namely, 1) short time subjective quality, denoted as STSQ, 2) playback indicator, denoted as PI, and 3) time elapsed since the last rebuffering event, denoted as $\textnormal{T}_\textnormal{R}$ \cite{NARX}.
The LSTM-QoE is evaluated on four publicly available continuous QoE databases and is shown to effectively capture the QoE dynamics with a high prediction performance.
To the best of our knowledge, this is the first work that performs a comprehensive evaluation over all publicly available continuous QoE databases and proposes an efficient QoE prediction method that delivers significantly higher performance compared to the state-of-the-art QoE evaluation methods.

The rest of the paper is organized as follows. Section \ref{sec:background} gives a brief overview of the existing QoE modeling approaches. The proposed LSTM-QoE model is presented in Section \ref{sec:QoE_modeling}.
The performance evaluation of the proposed model in explained in Section \ref{sec:perf_eval} along with a discussion on the  comparison with the existing QoE models. Finally, Section \ref{sec:conclusions} provides the concluding remarks.

\section{Background}
\label{sec:background}

QoE centric design has gained a lot of importance owing to several advantages it offers to multimedia service providers. Formulating descriptors and/or prediction models that quantify the end user QoE has been receiving enormous attention lately \cite{WeberFechner, IQX, Ricky, Singh_QoE, Balachandran2, Pessemier, Tobias, Deepti_HW, eTVSQ, Wang_ICIP2016}.
Real time multimedia applications such as online video streaming demand maintenance of an acceptable level of user QoE despite varying network conditions.
For providing a satisfactory quality of service, the end user QoE should be constantly monitored. 
The continuous monitoring of user QoE can enable network operators to optimize the utilization of network resources and stream videos to provide enhanced QoE.
Measuring the continuous QoE is a challenging task as it is highly subjective and dynamic in nature. However, many subjective evaluation studies have shown that although the preferences of individual subjects vary, by and large the QoE of users concur to a particular trend \cite{Ricky,Tobias,TVSQ_Chen,LFOVIA_QoE,LIVE_Netflix}. 
Subjective studies help a great deal in understanding the QoE and thereby facilitate the development of objective algorithms for quantifying the QoE.

Video quality assessment (VQA) plays a crucial role in the QoE prediction system \cite{LFOVIA_QoE}, \cite{NARX}, \cite{TVSQ_Chen}. VQA has been studied in several works in the literature \cite{MOVIE,STMAD,STRRED,BLIINDS,VIIDEO,FLOSIM,FLOSIM_NR}. A survey on the evolution of VQA measures is discussed in \cite{Winkler}. 
%While most of the VQA metrics are either full reference or reduced reference, few of them are no reference.
\cite{Chikkerur} provides a comprehensive study of various VQA metrics and suggests that the metrics MS-SSIM \cite{MSSSIM} and MOVIE \cite{MOVIE} provide a good video quality prediction performance. An optical flow based VQA method proposed in \cite{FLOSIM} is shown to provide a superior video quality prediction performance over all the existing methods. Even though the VQA metrics incorporate the aspects that determine user's perceptual quality, they are insufficient for determining streaming QoE \cite{LIVE_Netflix}. It is shown that the QoE is determined not just by the video quality but by a combination of factors such as rate adaptation and rebuffering events occurring at different time instants in a video session \cite{LFOVIA_QoE}, \cite{LIVE_Netflix}. Rate adaptation causes the video quality to fluctuate over time because of which the user QoE becomes time varying.

In wireless networks, the data rate delivered to the video user keeps varying with time due to channel fluctuations, mobility of the user, resource sharing etc. DASH allows the client to adapt its video rate to `best' match the data rate of the client. In spite of the best efforts, when the network/channel conditions degrade, the video client can run out of the playback content causing the playback to stall. Hence, the rate adaptation together with rebuffering events lead to a degradation of the user QoE.

There have been several efforts that address the challenge of QoE prediction for internet video delivery \cite{Balachandran1}, \cite{Balachandran2}.
The metrics that are defined in the 3GPP DASH specification TS 26.247 for QoE measurement have been identified in \cite{Singh_QoE}. Some of these include the average throughput, initial playout delay, buffer level etc. However, these metrics can only act as indicators of the QoE and cannot measure the actual QoE as they do not capture the perceptual experience of the user.
Other factors such as the initial loading time and startup delay have also been identified as the QoE influencing factors  \cite{Ricky}, \cite{Pessemier}, \cite{Tobias}, \cite{LIVE_Mobile_Stall_I}.
However, it is shown in these studies that shorter startup delays have minimal or almost negligible effect on the QoE. This suggests that the users are willing to wait for a considerable amount of time before the playback begins if they can be provided with a higher QoE. However, once the playback is started, the QoE of a user has been observed to be sensitive to time varying video quality as well as interruptions in the playback. Further, it is consistently observed in many QoE studies such as \cite{Ricky}, \cite{Pessemier}, \cite{Tobias}, \cite{LFOVIA_QoE} that the rebuffering events degrade the QoE severely. It is reported in \cite{Pessemier} that the users are willing to sacrifice higher resolution (or equivalently higher visual quality) for avoiding interruptions in the playback. Hence, it is imperative for the video client to maintain sufficient content in the playback buffer in order to avoid severe QoE degradation.

In \cite{TVSQ_Chen}, a Hammerstein-Wiener (HW) model has been proposed for measuring the time varying video quality due to rate adaptation. 
\cite{LIVE_Netflix} presents the LIVE Netflix Database along with a subjective study of the user QoE in the presence of both time varying quality and rebuffering events. Upon this database, a nonlinear autoregressive model is proposed using a neural network to predict the continuous QoE \cite{NARX}.
Similar to \cite{LIVE_Netflix}, the LFOVIA QoE database is presented in \cite{LFOVIA_QoE} along with a subjective study of continuous QoE of videos at Full-HD (FHD) and Ultra-HD (UHD) resolutions.
\cite{LFOVIA_QoE} also investigates the degradation in the continuous QoE when the viewers are subjected to various patterns of rate adaptation and rebuffering. 
Further, a continuous QoE prediction model is proposed based on Support Vector Regression (SVR).
In \cite{NLSS_QoE}, a QoE prediction model based on nonlinear state space (NLSS-QoE) has been proposed.
\cite{Deepti_contQoE} presents the Time-Varying QoE (TV-QoE) Indexer for predicting the continuous QoE using multi-stage and multi-learner HW approaches. These approaches employ multiple HW systems for modeling nonlinearity and memory effects and subsequently fuse their predictions for predicting the continuous time QoE.

Most QoE models proposed so far are evaluated and validated only on the QoE database for which they are designed.
To the best of our knowledge, there is no single QoE model that performs consistently well in a comprehensive evaluation across all available continuous QoE databases. 
Therefore, in this paper, 
we present LSTM-QoE, a QoE prediction model based on Long Short-Term Memory (LSTM) networks.
We demonstrate that the proposed model provides superior performance over the state-of-the-art QoE models on all the considered continuous QoE databases.
The proposed model is presented in the following section.

\section{QoE Modeling}
\label{sec:QoE_modeling}

\begin{figure}
\centering
\includegraphics[scale=0.85]{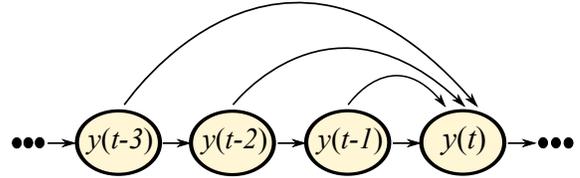}
\caption{Non-Markovian nature of QoE process with long-term dependencies.}
\label{fig:Markov}
\end{figure}

The user QoE in video streaming is determined by the human visual perception \cite{recency}. According to International Telecommunications Union, QoE is defined as the overall quality of an application or a service as \textit{perceived subjectively} by the end user \cite{ITU_QoE}. 
Many psychovisual experimental studies have hypothesized that the relationship between the visual quality and the perceptual experience in the human visual system (HVS) is highly nonlinear in nature \cite{albrecht}, \cite{Kalpana}. This is due to the nonlinear response properties of the neurons in the primary visual cortex. Because of this, the QoE behavioral patterns of a user while watching a video are nonlinear functions of the stimulus. Further, it has been observed through subjective studies that the visual QoE in general is dynamic and time varying in nature, varying continuously in response to a series of QoE influencing events such as rebuffering \cite{Ricky} and rate adaptation \cite{TVSQ_Chen}. 
Due to these events, the HVS produces hysteresis effect \cite{Kalpana_hysteresis}, wherein, the past event occurrences leave a considerable impact on the QoE at the current time instant. 
This is particularly observed to be prominent in the cases where the effects of poor visual quality segments occurring in the past ripple through and produce a significant impact on the current QoE, even though the visual quality rendered at the current instant is higher. 
Hysteresis effect essentially implies that the QoE process is non-Markovian in nature, as there exists a memory of a sequence of events beyond first order influencing the current QoE.
The QoE process can have time varying long and short-term dependencies as it evolves continuously with time.
Such dependencies can be modeled using a higher order process as illustrated in Fig. \ref{fig:Markov},
where the current QoE has influences from the previous QoE values.

In summary, the continuous QoE is a nonlinear stochastic process exhibiting non-Markovian temporal dynamics
due to the hysteresis effect. 
To capture such dynamics, we employ LSTMs, a class of recurrent neural networks
that has been shown to be effective in modeling 
sequential data having long/short-term dependencies \cite{LSTM, LSTM_2000}. LSTMs have been successfully used to address complex challenges in applications such as sequence labeling \cite{sequence_labeling}, visual recognition \cite{LSTM_recognition}, image captioning \cite{LSTM_imagecaption} and machine translation \cite{RNN_NLP}.
Hence, in the following subsection, we propose an LSTM based approach to model the continuous QoE prediction.

\subsection{LSTM-QoE} 

\begin{figure}
\centering
\includegraphics[scale=0.50]{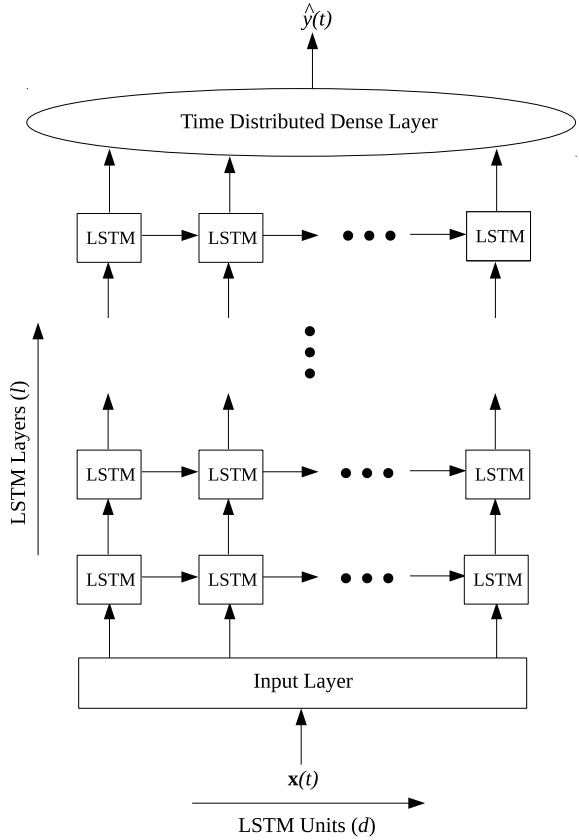}
\caption{Proposed LSTM network for QoE prediction.}
\label{fig:LSTM_architecture}
\end{figure}

Let the actual and the predicted QoE at time instant $t$ be represented by $y(t)$ and $\hat{y}(t)$, respectively. Let $\textbf{x}(t) \in \mathbb{R}^m_{\geq0}$ represent the feature set that takes values from a $m$-dimensional space of non-negative real numbers. Here, the feature set $\textbf{x}(t)$ is representative of the QoE influencing (determining) factors that govern the QoE evolution. Thus, at any given time instant $t$, we use the time-indexed feature vector $\textbf{x}(t) = [x_1(t) \ x_2(t) \ \cdots \ x_m(t)]$ 
%to capture the current QoE influences so as 
to predict the current QoE $\hat{y}(t)$.
Given that the QoE $y(t)$ is non-Markovian, we have the following \cite{SMC_NIPS}:
\begin{equation}
p(y(t)|y(t-1),y(t-2),\cdots,y(1)) \neq p(y(t)|y(t-1)), \nonumber
\label{eq:LSTM_condprob}
\end{equation}
where, the conditional probability $p(y(t)|y(t-1),y(t-2),\ldots,y(1))$ indicates that the QoE involves 
higher order temporal dependencies. 
These dependencies could be complex and it may not be possible for a single LSTM to effectively capture them 
\cite{LSTM_recognition}.
Hence, we propose a network of LSTMs to learn these dependencies involved in the QoE process as depicted in Fig. \ref{fig:LSTM_architecture}. 
The motivation for this proposal comes from various LSTM based solutions that have been shown to be successful in addressing some of the problems involving complex dependencies such as sequence to sequence learning \cite{LSTM_Sutskever}, activity recognition and image description \cite{LSTM_recognition}.

Fig. \ref{fig:LSTM_architecture} shows the proposed multi-layered multi-unit LSTM network for QoE prediction. 
The proposed network is a cascade of several LSTM units that are stacked up to constitute LSTM layers. Let the proposed LSTM network be denoted by $LSTM_{l,d}$, where $l$ represents the number of layers and $d$ the number of units in each layer. The parameters $l$ and $d$ are the design choices that are to be tuned based on the nature of the underlying process and the complexity of the dependencies.
Using the input features $\textbf{x}(t)$, the LSTM network %learns the dependencies and 
computes the QoE estimate $\hat{y}(t)$ continuously at every time instant $t$. 
Each LSTM unit tracks the stochastic process by maintaining an internal cell state, referred to as latent state, and the state transitions are driven by the input features $\textbf{x}(t)$. Let $\textbf{c}(t)$ represent the set of LSTM cell states in the network.
LSTMs are modeled to learn the underlying complex distribution governing the state transition and predict the QoE at every time instant as follows \cite{SMC_NIPS}, \cite{SSL}:
\begin{equation}
p(y(t)|y(t-1),y(t-2),\cdots,y(1)) = p(y(t);g(\textbf{c}(t))), \nonumber
\label{eq:LSTM_1}
\end{equation}
where, $g(\cdot)$ refers to a differentiable function that maps $\textbf{c}(t)$ of LSTMs to the parameters of the underlying unknown QoE distribution. $LSTM_{l,d}$ provides two functionalities: 1) $LSTM_{l,d}^o$ for the output QoE prediction and 2) $LSTM_{l,d}^c$ for the cell state update.
The predicted QoE $\hat{y}(t)$ is given by
\begin{equation}
\hat{y}(t) = LSTM_{l,d}^o(\textbf{x}(t),\textbf{c}(t-1)).
\end{equation}

The cell state update \cite{SSL}
for the LSTM network is given by
\begin{equation}
\textbf{c}(t) = LSTM_{l,d}^c(\textbf{c}(1:t-1),\hat{y}(1:t-1)), \forall t > 1.
\label{eq:LSTM_2}
\end{equation}
The cell state $\textbf{c}(t)$ is a deterministic function of the past QoE $\hat{y}(1:t-1)$ and the past cell states through the LSTM network function $LSTM_{l,d}^c$. 
This enables the state vector $\textbf{c}(t)$ to track complex temporal dependencies in the QoE process and thereby empower LSTMs to model the sequential data.
The predicted QoE $\hat{y}(t)$ is obtained using the current input feature $\textbf{x}(t)$ and the cell state before the update $\textbf{c}(t-1)$ as provided in (\ref{eq:LSTM_1}).
Further, the nonlinearities involved in the QoE prediction are also taken into account, as LSTMs inherently possess nonlinearity by construction \cite{LSTM}. 

From (\ref{eq:LSTM_1}) and (\ref{eq:LSTM_2}), we observe that the selection of input features $\textbf{x}(t)$ is crucial for continuous QoE prediction. The selected input features should be such that they effectively capture and integrate various influences governing QoE evolution through the LSTM states. Hence, we discuss the constitution of the input feature vector $\textbf{x}(t)$ in the following subsection.

\subsection{Feature Selection}

Due to their demonstrated efficiency, we employ the following three features for QoE prediction in the proposed LSTM-QoE \cite{NARX}, \cite{NLSS_QoE}:

\begin{itemize}

\item[1)] \textit{Short Time Subjective Quality} (STSQ): STSQ refers to the perceptual quality of the video segment currently being rendered to the user.
STSQ can be measured using off-the-shelf video quality assessment (VQA) metrics such as STRRED \cite{STRRED}, MS-SSIM \cite{MSSSIM}. STSQ as a feature has been successfully employed for QoE prediction in  \cite{NARX} and \cite{NLSS_QoE}.

\item[2)] \textit{Playback Indicator} (PI): A binary indicator variable PI to indicate whether the video is currently in the playback state or in the rebuffering state. The playback status indicator has been shown to be effective for QoE prediction in the earlier approaches such as NARX \cite{NARX} and SVR-QoE \cite{LFOVIA_QoE}.

\item[3)] \textit{Time elapsed since last rebuffering} ($\textnormal{T}_\textnormal{R}$): 
Since, the user QoE is heavily influenced by the occurrence of rebuffering events, we employ $\textnormal{T}_\textnormal{R}$, a variable to keep track of the time elapsed since the occurrence of the last rebuffering event. $\textnormal{T}_\textnormal{R}$ has been used for QoE prediction in \cite{NLSS_QoE}.

\end{itemize}

We subsequently show that the proposed model driven by these features is powerful enough to provide superior prediction that significantly outperforms the state-of-the-art QoE prediction models.
We discuss the performance evaluation of LSTM-QoE over continuous QoE databases in the following section.

\section{Performance Evaluation of LSTM-QoE}
\label{sec:perf_eval}

In this section, we consider the performance evaluation of the proposed model over continuous QoE databases. We first describe the databases and the evaluation procedure, followed by a discussion on the performance measures. We then explain the selection of the parameters $l$ and $d$ in the proposed LSTM-QoE network. Further, we discuss the efficacies of the individual features for QoE prediction. Using the best network configuration and the best performing feature set, we present the performance evaluation of the proposed model for QoE prediction.

\subsection{Database Description}
\label{subsec:QoE_prediction}

We employ four publicly available continuous QoE databases for evaluation. The details of these databases along with the training and the test procedures followed in the study are described as follows. \\

\begin{itemize}

\item[1)]  LIVE Netflix Database \cite{LIVE_Netflix}: The database provides 112 videos of which 56 videos have compression (encoding) artifacts only and the remaining 56 videos have compression artifacts and rebuffering combined together. In this database, 112 videos are constructed out of 14 videos that are distinct in content with 8 videos per content. Each of these 8 videos has a unique playout pattern. 
The videos in the database have a resolution of 1920$\times$1080.
The continuous QoE scores of the videos in the database have a dynamic range of [-2.26, 1.52]. Lower the score values, lower is the QoE. \\

We employ the standard training and test procedure with a training-test split as described in \cite{NARX}.
Accordingly, in each training-test split, one video in the database is considered in each test set. The model is trained on the set of videos that do not have the same content and the playout pattern as those of the video in the test set. 
This excludes 21 videos (14 with the same playout pattern and 7 with the same content) from the training process for each test video.
Thus, in each training-test split, the training set consists of 91 videos out of a total of 112 videos in the database. This procedure is employed in order to ensure a fair evaluation of the trained model.
This process is carried out for all the videos in the database. 
Hence, there are 112 test evaluations corresponding to each of the videos.  \\

\item[2)] LFOVIA QoE Database \cite{LFOVIA_QoE}: The database consists of 36 distorted videos 
derived from 18 reference videos, each having a duration of 120 seconds. These 36 videos 
contain a combination of time varying quality and rebuffering distortions. In this database, 18 of the 36 videos are at full-HD (1920$\times$1080) resolution, while the other 18 videos are at ultra-HD (3840$\times$2160) resolution. The QoE scores obtained for the videos in the database are in the range [0, 100], with score 0 being the worst and 100 being the best. \\

We employ a training-test split procedure similar to that employed for the LIVE Netflix Database, wherein, there is only one video in each test set. The training set is constituted by the videos that do not contain the playout pattern as that of the video in the test set. In other words, the videos in the database having the same playout pattern as that of the test video are excluded from training.
Accordingly, 25 of 36 videos are chosen for training the model for each test video.
Thus, there are 36 test evaluations corresponding to each of the videos in the database.

\item[3)]  LIVE QoE Database \cite{TVSQ_Chen}: The database consists of 15 time varying quality videos generated from 3 pristine reference videos. Each video is of length 300 seconds with a resolution 1280$\times$720. The QoE scores obtained for the videos in the database
are in the range [0, 100], with score 0 being the worst and 100 being the best. \\

For the LIVE QoE Database, we employ the leave $p$-out cross validation methodology for performance evaluation, with a value of $p$ = 5. A similar methodology for evaluation has also been employed in \cite{TVSQ_Chen} and \cite{C3D_TVSQ}. Accordingly, for the evaluation of each test video, the training is performed using those 10 videos that differ in the video content as well as the time varying quality pattern as compared to those of the test video. Accordingly, 10 out of 15 videos in the database are employed for training in each training-test split. 
The videos in this database have only time varying quality artifacts and no rebufferings.
Hence, we employ STSQ as the only input feature in the proposed model for this database. A similar setting has been employed for QoE prediction in \cite{Deepti_contQoE}.
Alternately, the feature PI can be set to 1 meaning `ON' and $\textnormal{T}_\textnormal{R}$ constant throughout the video duration in our proposed model for this database. \\

\item[4)] LIVE Mobile Video Stall Database-II \cite{LIVE_Mobile_Stall_II}: The database consists of 174 videos with rebuffering events occurring at various locations in the video playback. The database investigates the continuous evaluation of QoE due to rebuffering events only. \\

In the LIVE Mobile Video Stall Database-II, the distortion patterns are randomly distributed across the videos. Hence, we employ a slightly different evaluation methodology on the lines similar to that employed for LIVE Netflix and LFOVIA QoE databases. Accordingly, we create 174 test sets corresponding to each of 174 videos in the database. For each test set, we randomly choose 80\% videos from the remaining 173 videos for training the model and perform evaluation over the test video.

\end{itemize}

The various measures employed for quantifying the QoE prediction performance are explained next.

\begin{figure*}[t]
\centering
\begin{subfigure}[b]{0.24\textwidth}
\includegraphics[scale=0.32]{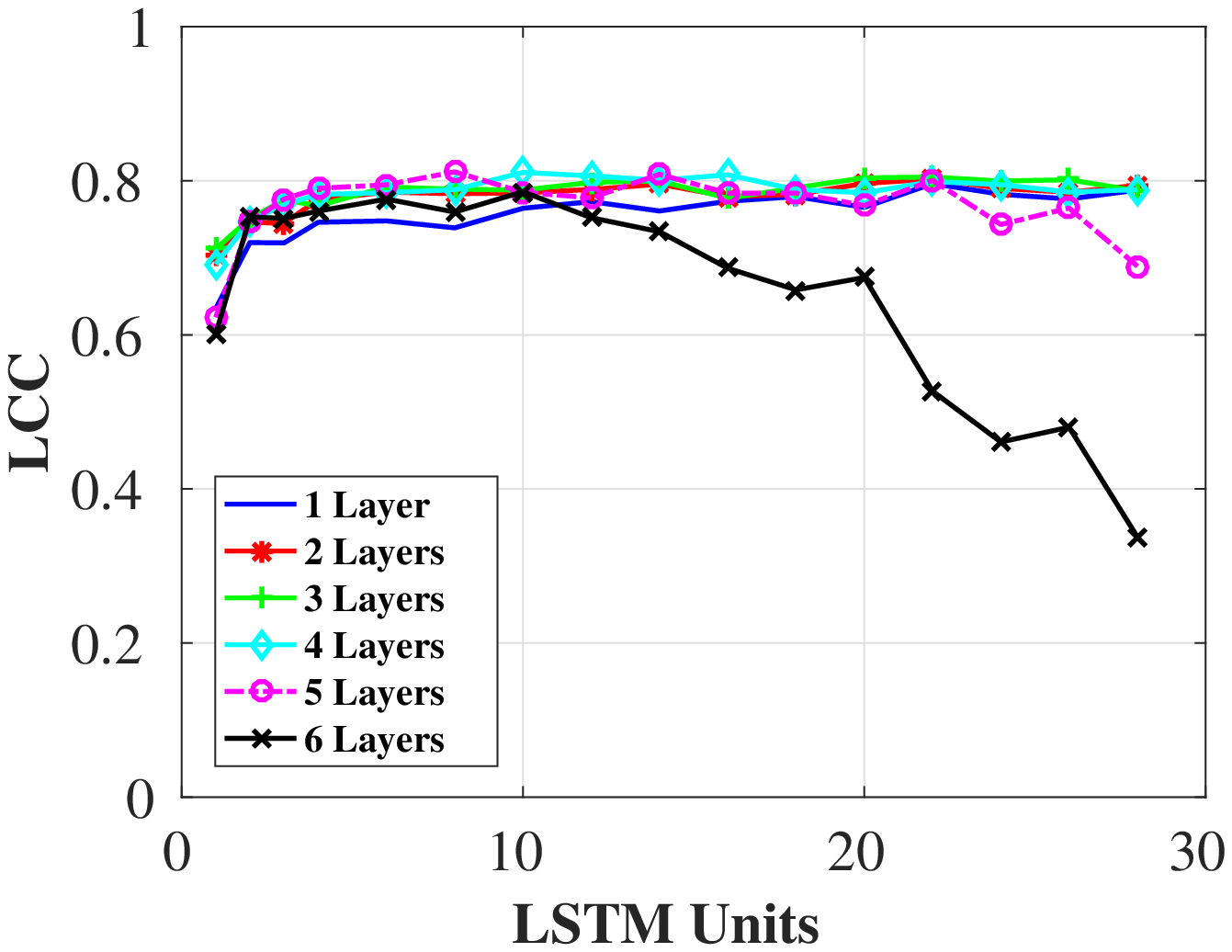}
\subcaption{LIVE Netflix \cite{LIVE_Netflix}: LCC.}
\label{subfig:LSTM_layers_LIVE_Netflix_LCC}
\end{subfigure}
\begin{subfigure}[b]{0.24\textwidth}
\includegraphics[scale=0.32]{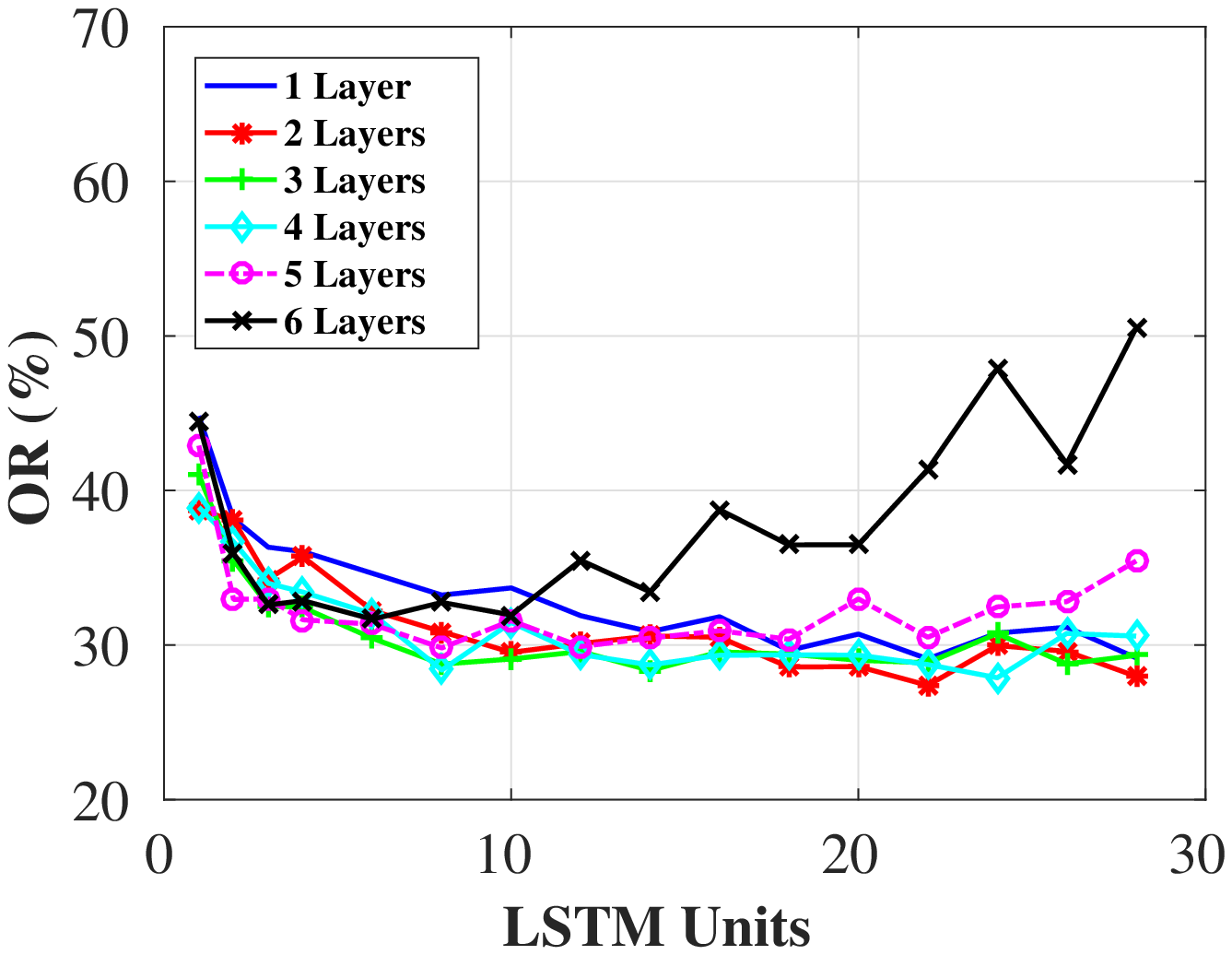}
\subcaption{LIVE Netflix \cite{LIVE_Netflix}: OR (\%).}
\label{subfig:LSTM_layers_LIVE_Netflix_OR}
\end{subfigure}
\begin{subfigure}[b]{0.24\textwidth}
\includegraphics[scale=0.32]{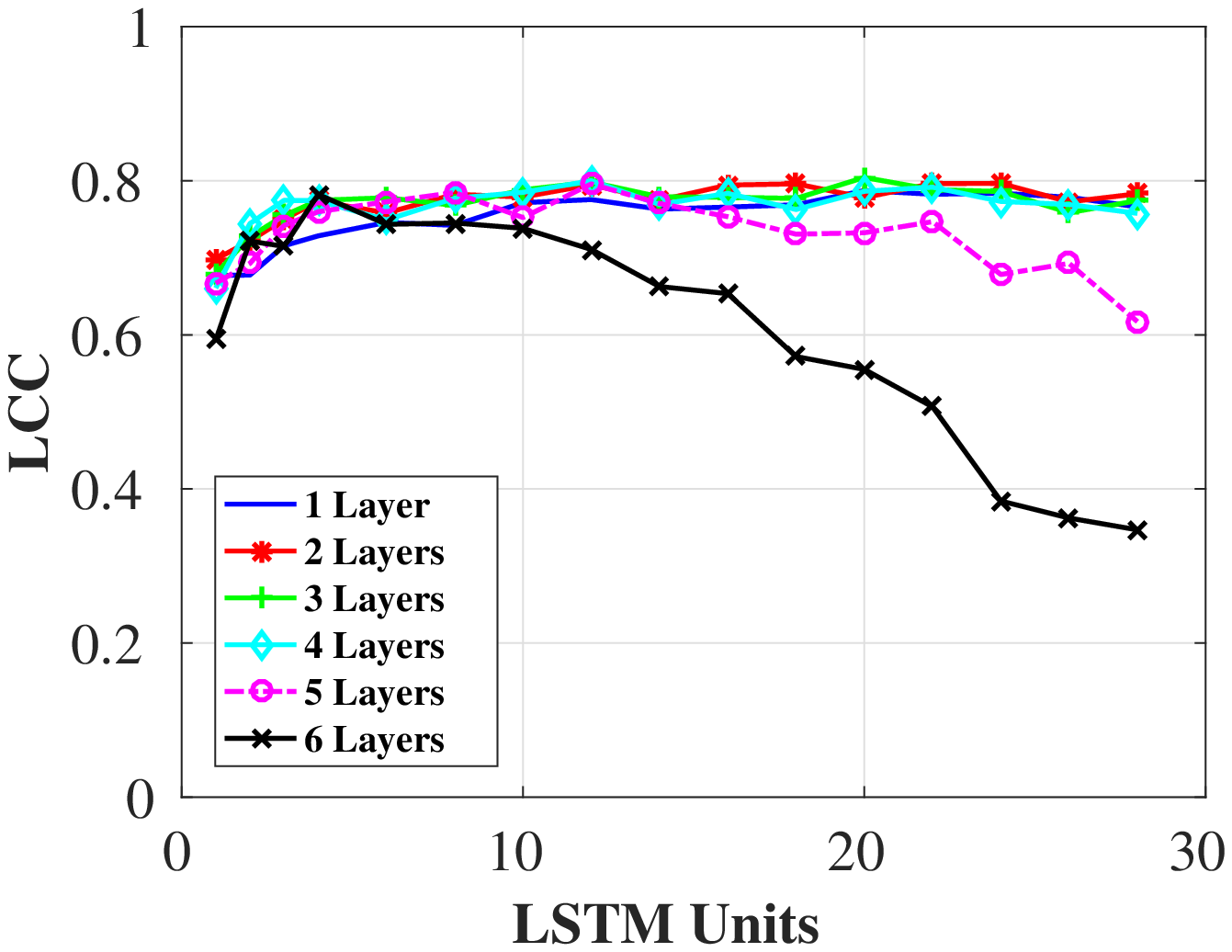}
\subcaption{LFOVIA QoE \cite{LFOVIA_QoE}: LCC.}
\label{subfig:LSTM_layers_LFOVIA_QoE_LCC}
\end{subfigure}
\begin{subfigure}[b]{0.24\textwidth}
\includegraphics[scale=0.32]{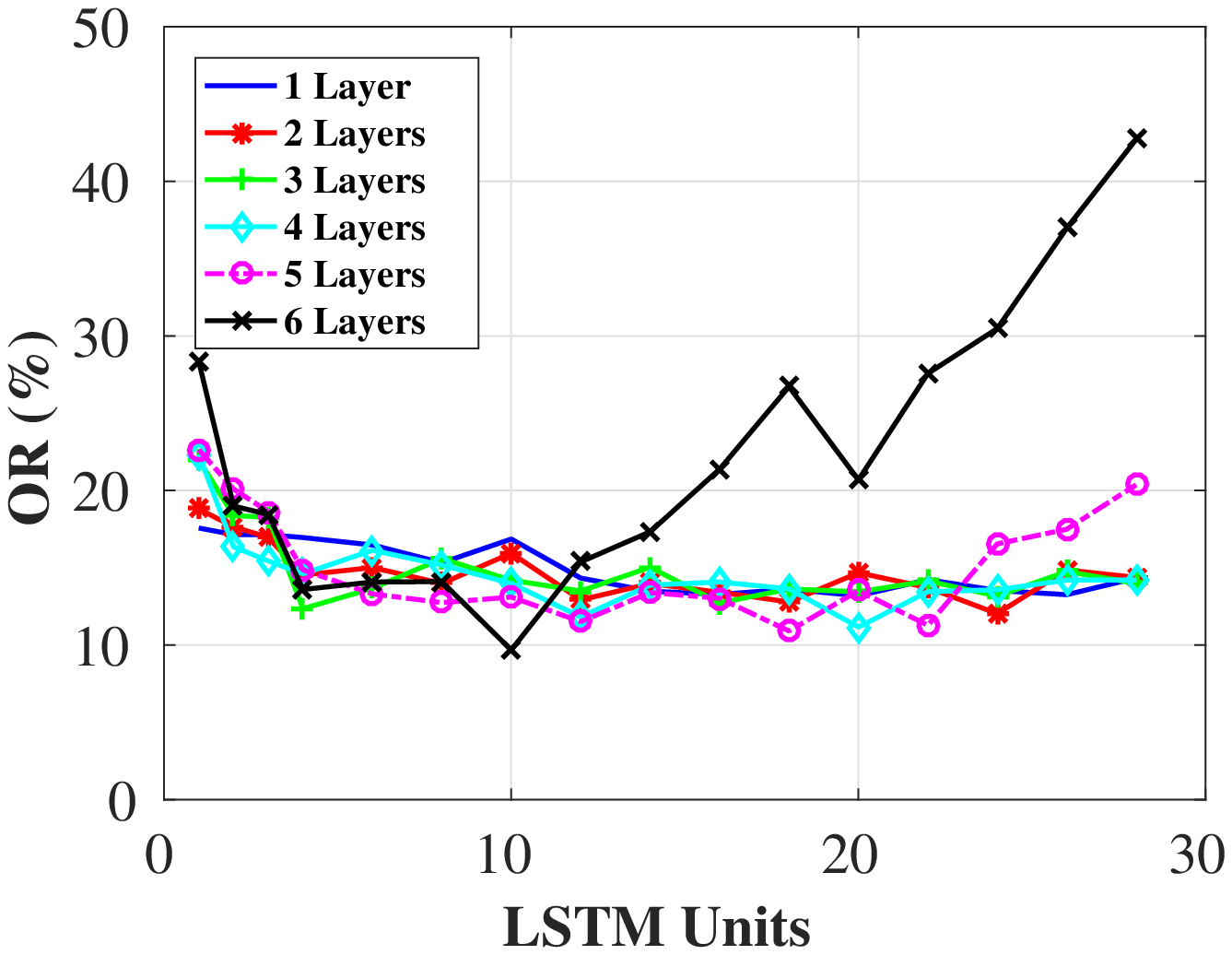}
\subcaption{LFOVIA QoE \cite{LFOVIA_QoE}: OR (\%).}
\label{subfig:LSTM_layers_LFOVIA_QoE_OR}
\end{subfigure}
\caption{QoE prediction performance of various LSTM-QoE network configurations.}
\small
Figs. \ref{subfig:LSTM_layers_LIVE_Netflix_LCC}, \ref{subfig:LSTM_layers_LIVE_Netflix_OR}, \ref{subfig:LSTM_layers_LFOVIA_QoE_LCC}, and \ref{subfig:LSTM_layers_LFOVIA_QoE_OR} illustrate the QoE prediction performance for different LSTM units and layers upon the LIVE Netflix \cite{LIVE_Netflix} and the LFOVIA QoE \cite{LFOVIA_QoE} Databases in terms of LCC and OR measures.
\label{fig:LSTM_layers}
\end{figure*}

\begin{figure*}[t]
\centering
\begin{subfigure}[b]{0.37\textwidth}
\includegraphics[scale=0.34]{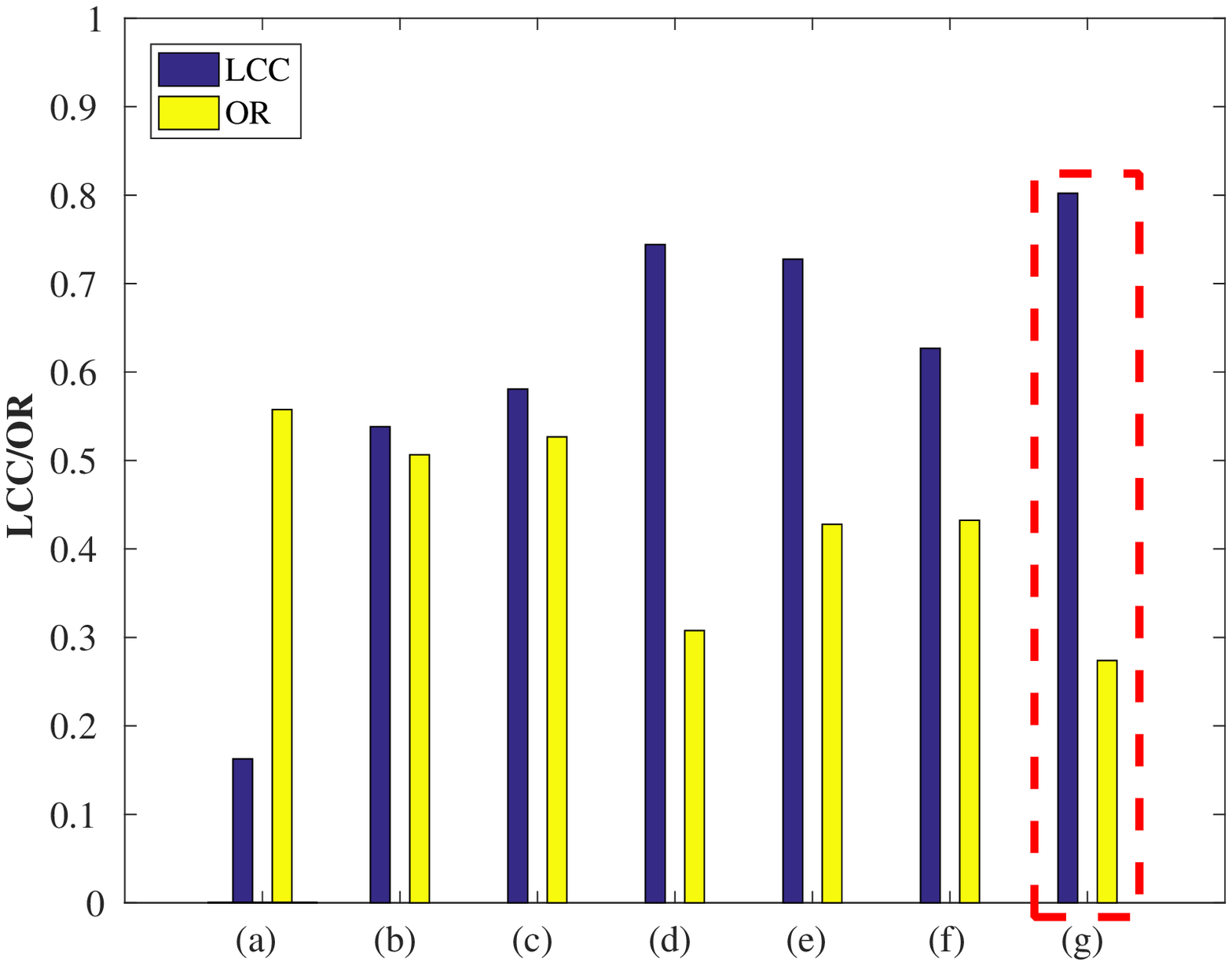}
\subcaption{LIVE Netflix Database \cite{LIVE_Netflix}.}
\label{subfig:LSTM_features_LIVE_Netflix}
\end{subfigure}
\begin{subfigure}[b]{0.37\textwidth}
\includegraphics[scale=0.34]{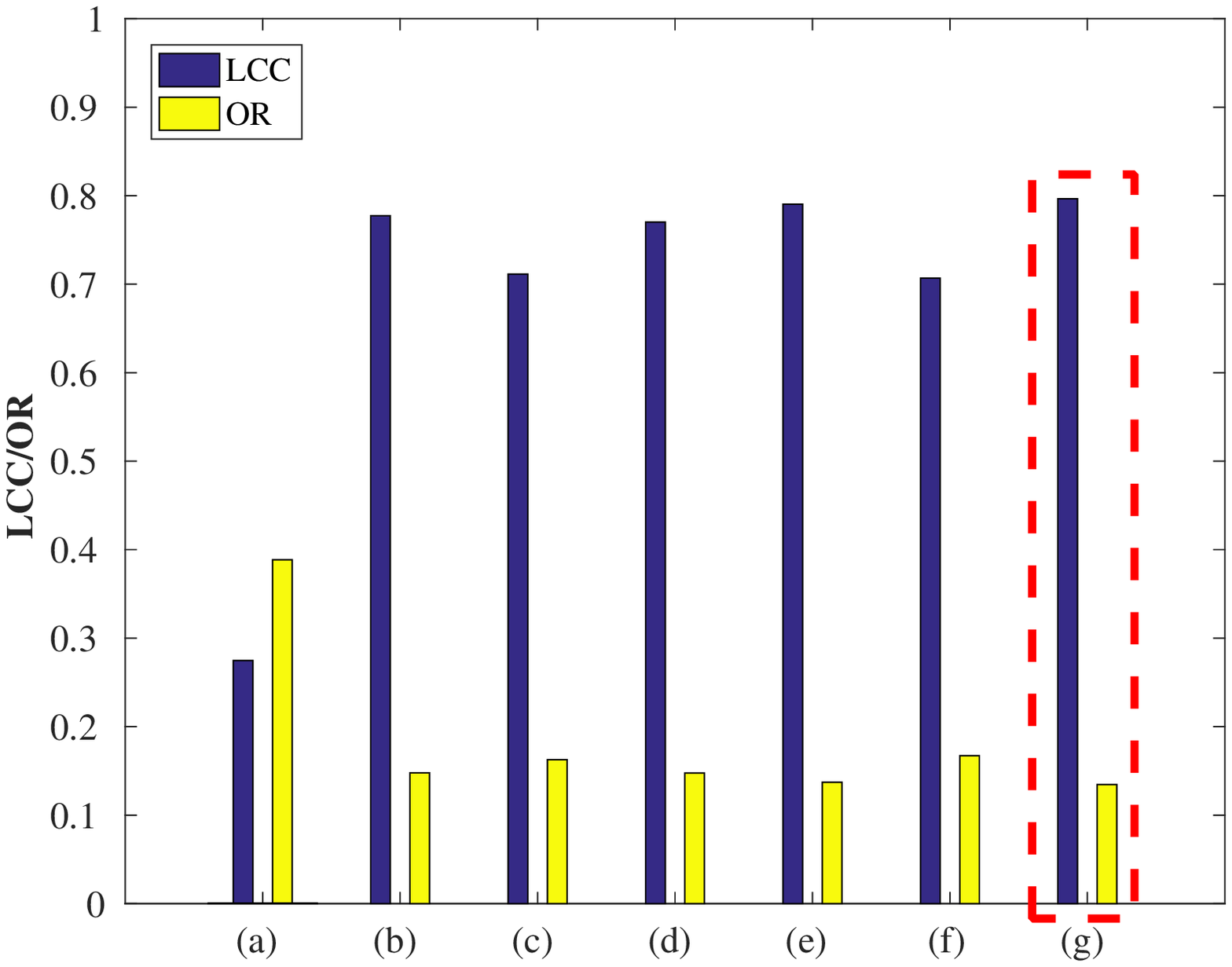}
\subcaption{LFOVIA QoE Database \cite{LFOVIA_QoE}.}
\label{subfig:LSTM_features_LFOVIA_QoE}
\end{subfigure}
\caption{QoE prediction performance of various feature combinations.}
\small
Figs. \ref{subfig:LSTM_features_LIVE_Netflix} and \ref{subfig:LSTM_features_LFOVIA_QoE} illustrate the QoE prediction performance for various feature combinations upon the LIVE Netflix \cite{LIVE_Netflix} and the LFOVIA QoE \cite{LFOVIA_QoE} Databases, respectively. The various feature combinations are as follows: (a) STSQ, (b) PI, (c) $\textnormal{T}_\textnormal{R}$, (d) STSQ+PI, (e) PI+$\textnormal{T}_\textnormal{R}$, (f) STSQ+$\textnormal{T}_\textnormal{R}$, and (g) STSQ+PI+$\textnormal{T}_\textnormal{R}$. Dotted red box indicates the best performing feature combination.
\label{fig:LSTM_features}
\end{figure*}

\subsection{Performance Evaluation Measures}

The performance of QoE prediction using the proposed model is quantified using the following four measures: 1) Linear Correlation Coefficient (LCC), 2) Spearman Rank Order Correlation Coefficient (SROCC), 3) Normalized Root Mean Squared Error ($\textnormal{RMSE}_\textnormal{n}$), and 4) Outage Rate (OR) \cite{TVSQ_Chen}, \cite{LFOVIA_QoE}.

The LCC and SROCC provide a quantification of the correlation between the predicted QoE scores  and the ground truth QoE scores. The $\textnormal{RMSE}_\textnormal{n}$ and OR measure the closeness between the predicted scores and the ground truth scores. The QoE databases considered in our evaluation have different QoE score ranges. Hence, we normalize the actual RMSE values to obtain the normalized RMSE `$\textnormal{RMSE}_\textnormal{n}$'. Since the predicted scores are continuous, it is insufficient to assess the performance using any of these metrics alone \cite{LFOVIA_QoE}, \cite{C3D_TVSQ}.
%, hence we compute all the performance measures to quantify completely. 
Therefore, we consider all the aforementioned measures, i.e., LCC/SROCC and $\textnormal{RMSE}_\textnormal{n}$/OR jointly to assess the performance of the QoE prediction model. 
A good performing model is characterized by high LCC/SROCC (closer to 1) and low $\textnormal{RMSE}_\textnormal{n}$/OR (closer to 0).
We next discuss the parameter selection for the LSTM-QoE network.

\subsection{Parameter Selection for LSTM-QoE Network}
\label{subsec:LSTM_layers_units}

\begin{figure*}[t]
\centering
\includegraphics[scale=0.25]{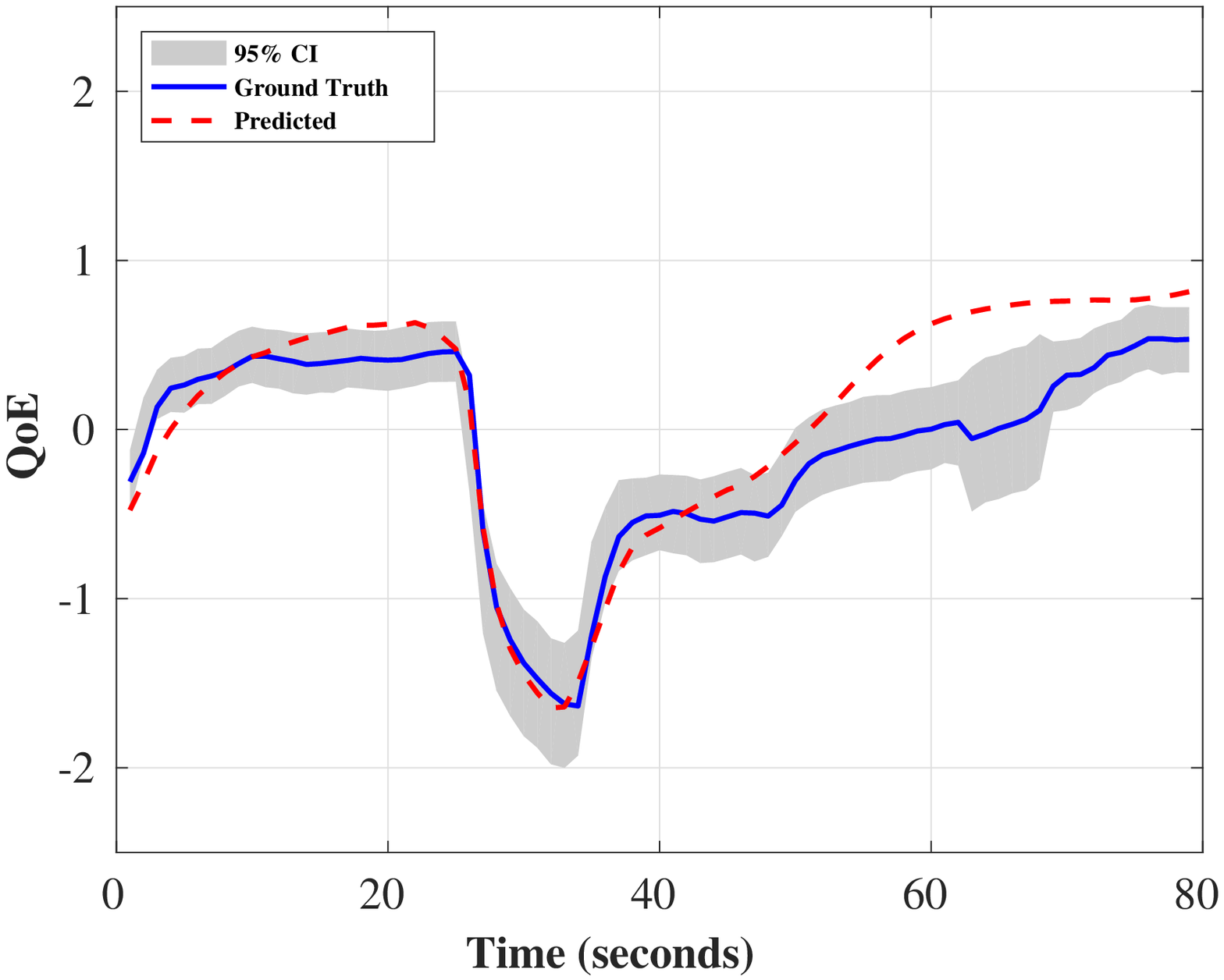}
\includegraphics[scale=0.25]{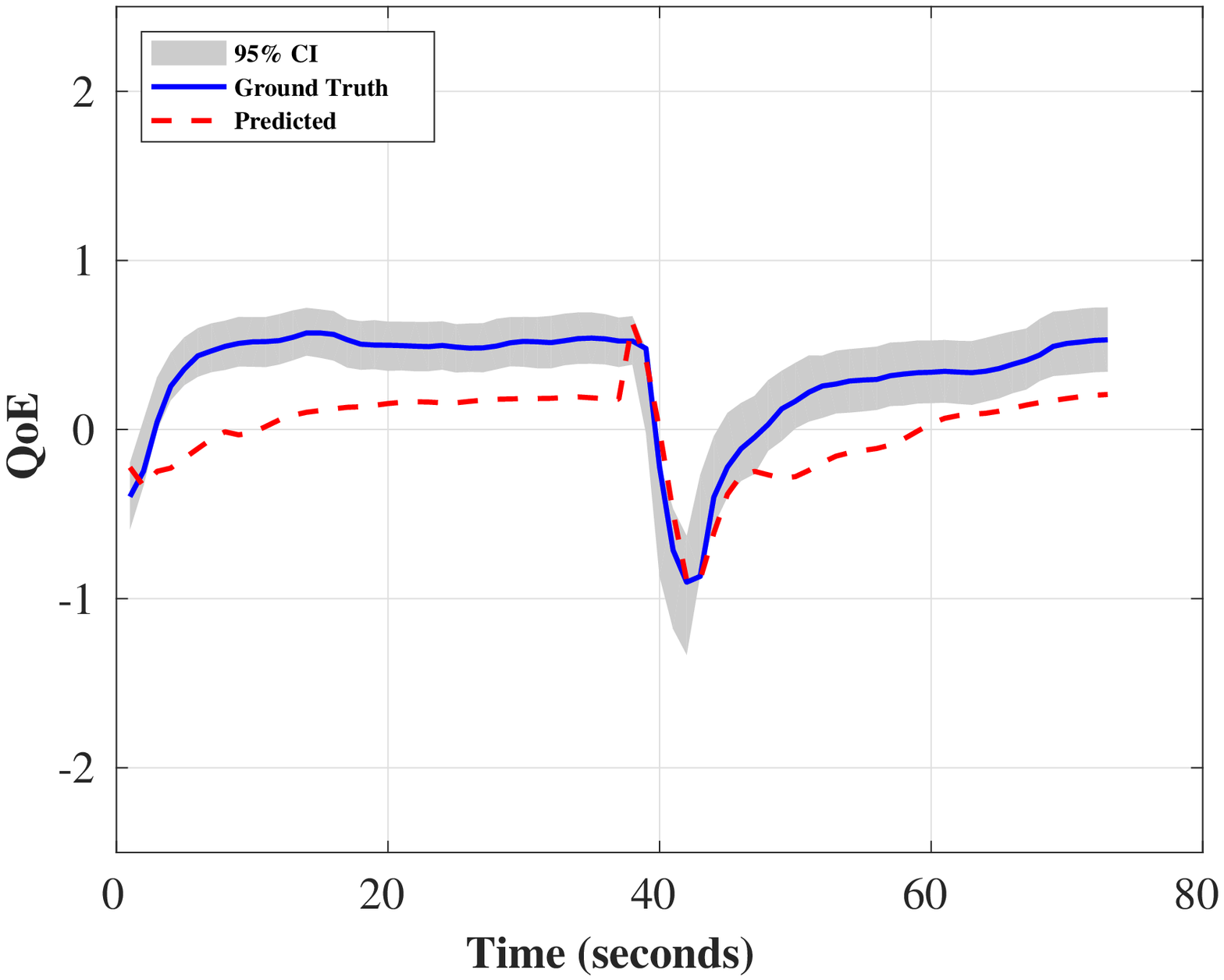}
\includegraphics[scale=0.25]{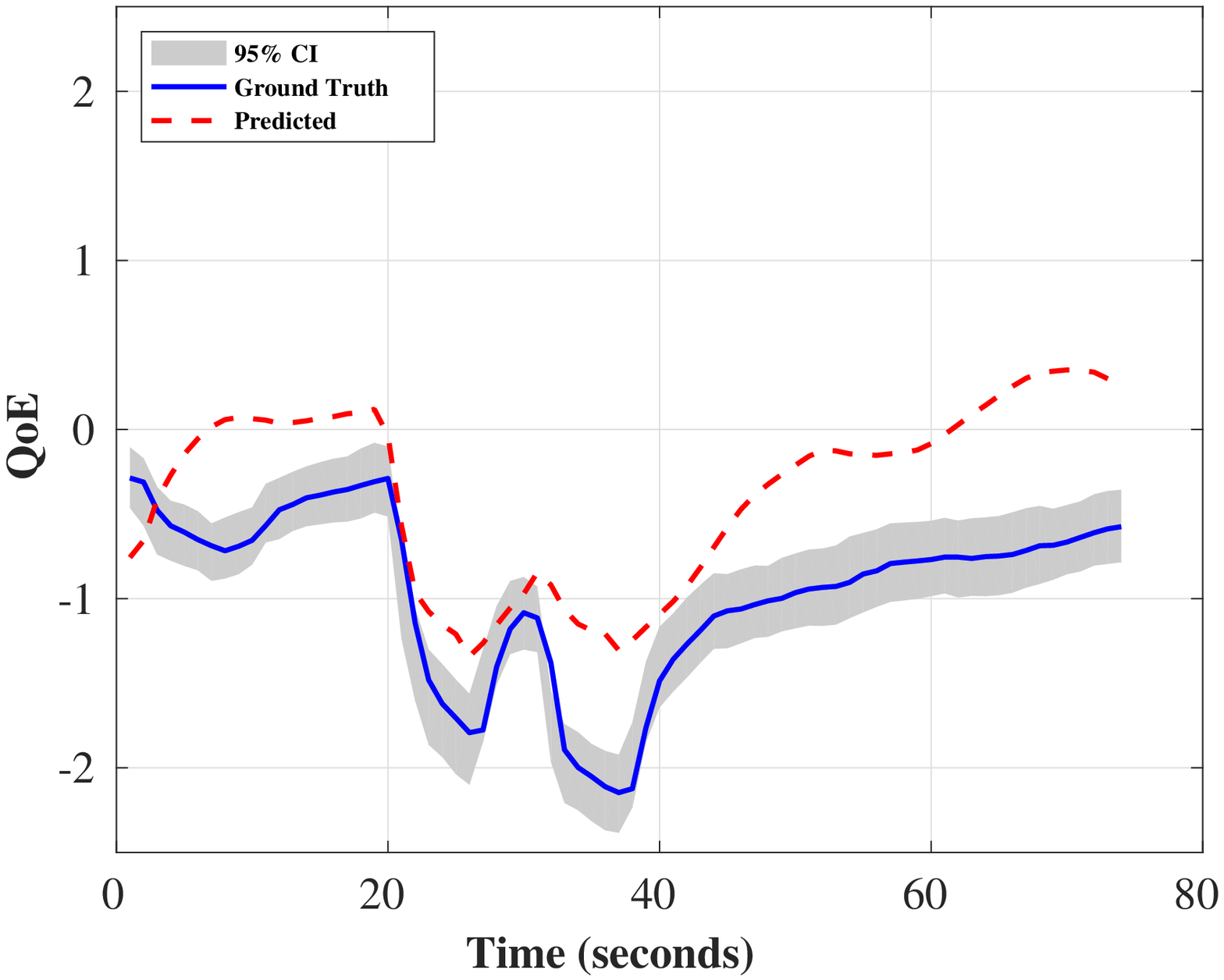}
\includegraphics[scale=0.25]{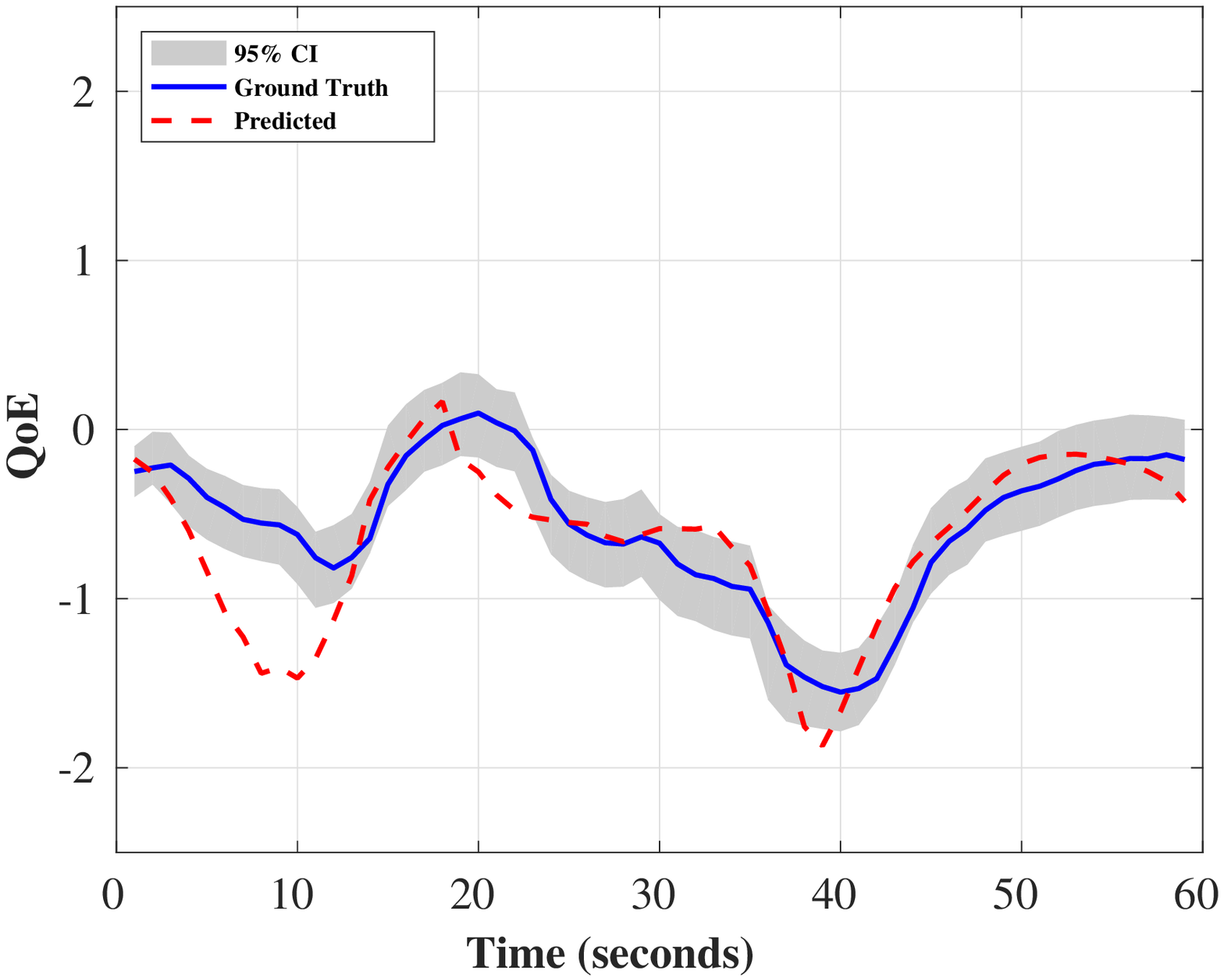}
\caption{QoE prediction performance of the LSTM-QoE model over the LIVE Netflix Database \cite{LIVE_Netflix} with STRRED as the STSQ metric.}
\label{fig:LIVE_Netflix_LSTM}
\end{figure*}

\begin{figure*}[t]
\centering
\includegraphics[scale=0.25]{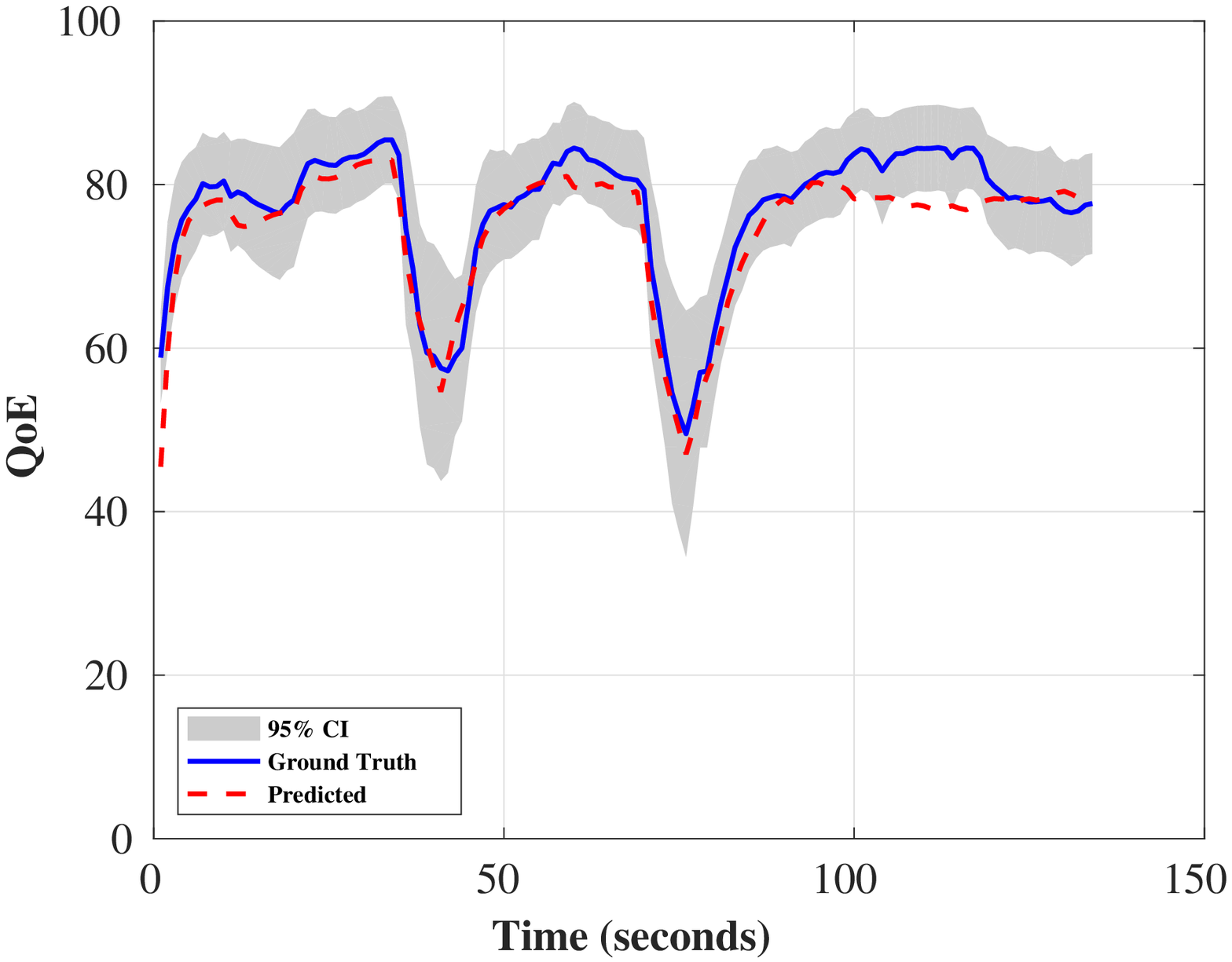}
\hspace{-0.3cm}
\includegraphics[scale=0.25]{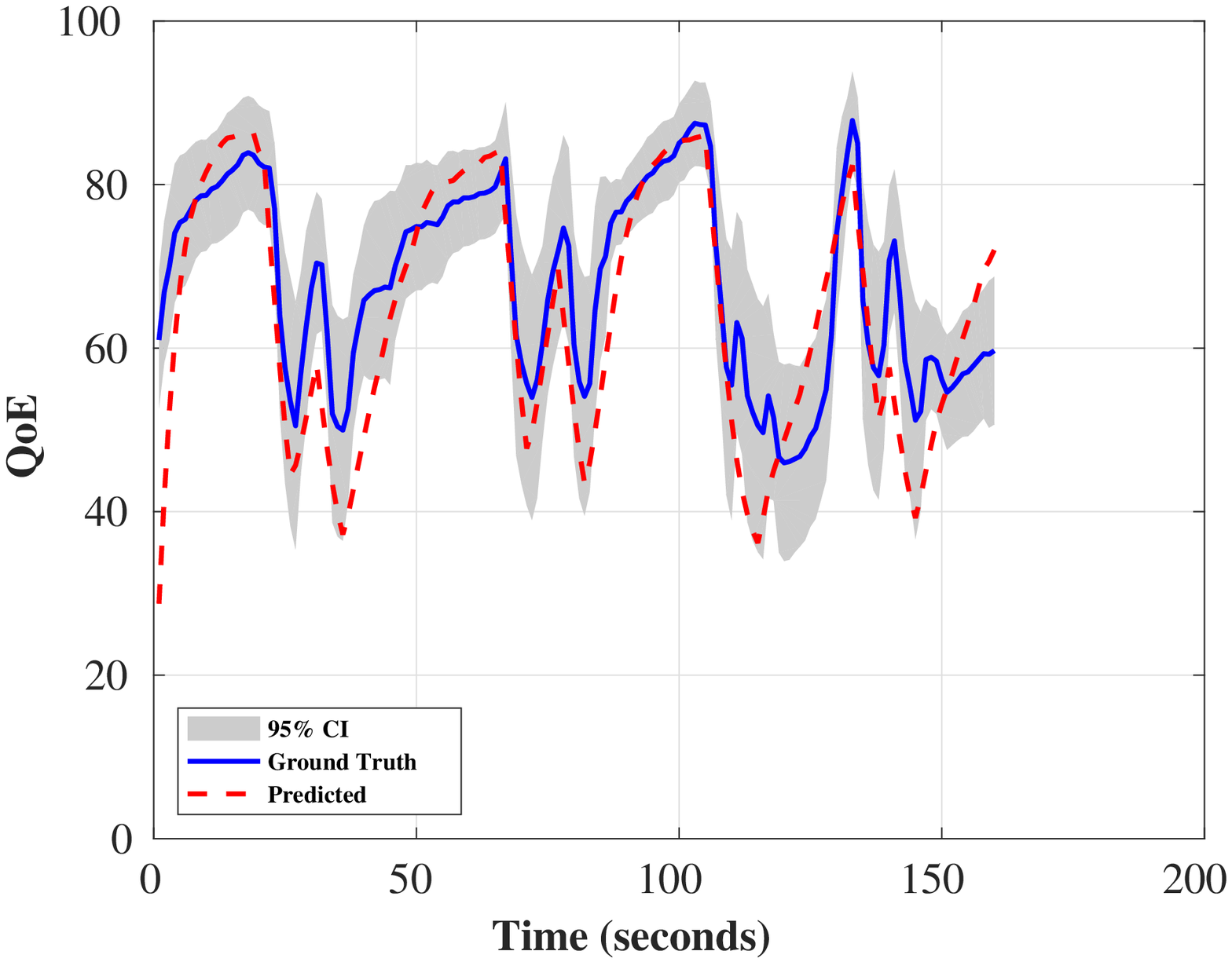}
\hspace{-0.3cm}
\includegraphics[scale=0.25]{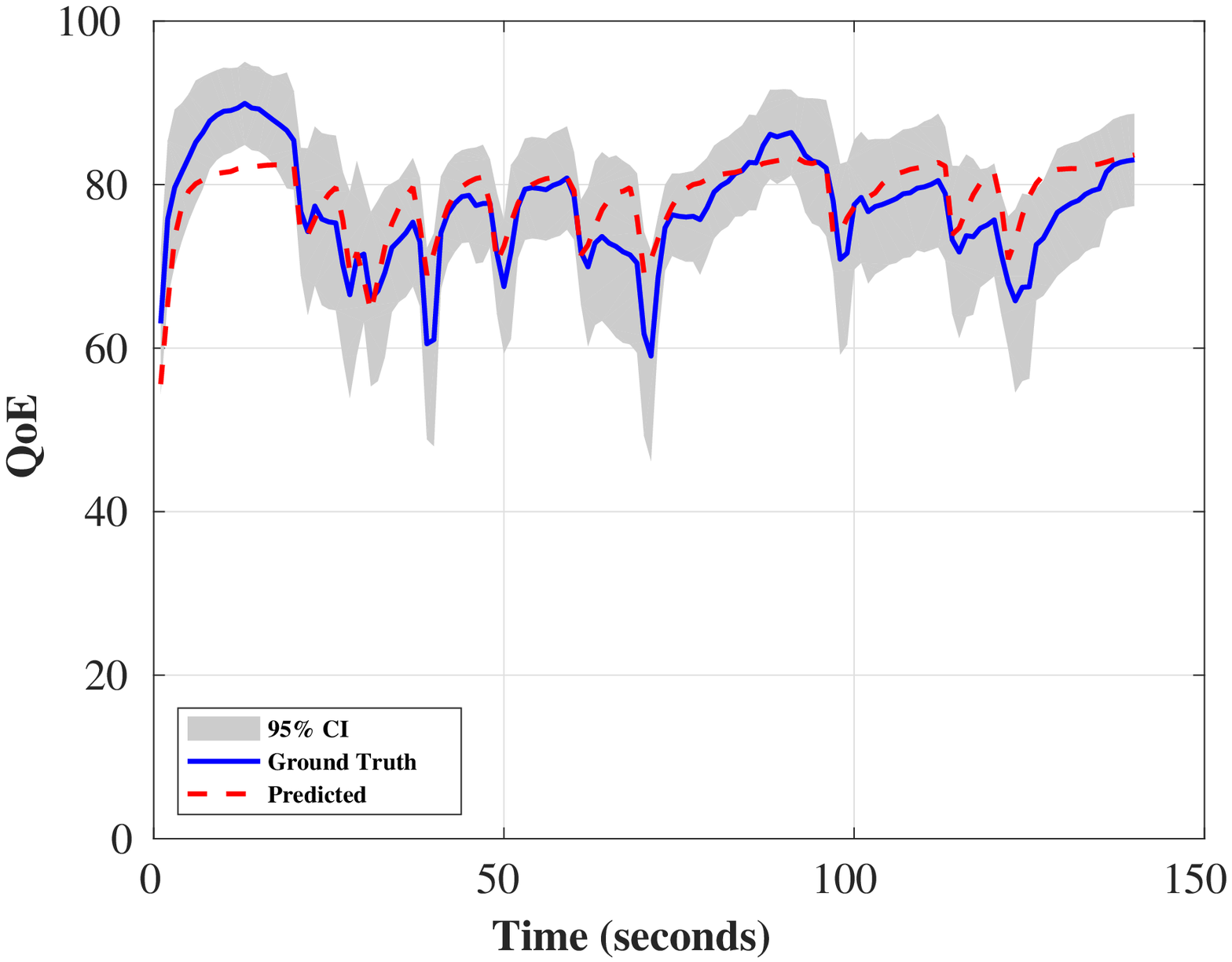}
\hspace{-0.3cm}
\includegraphics[scale=0.25]{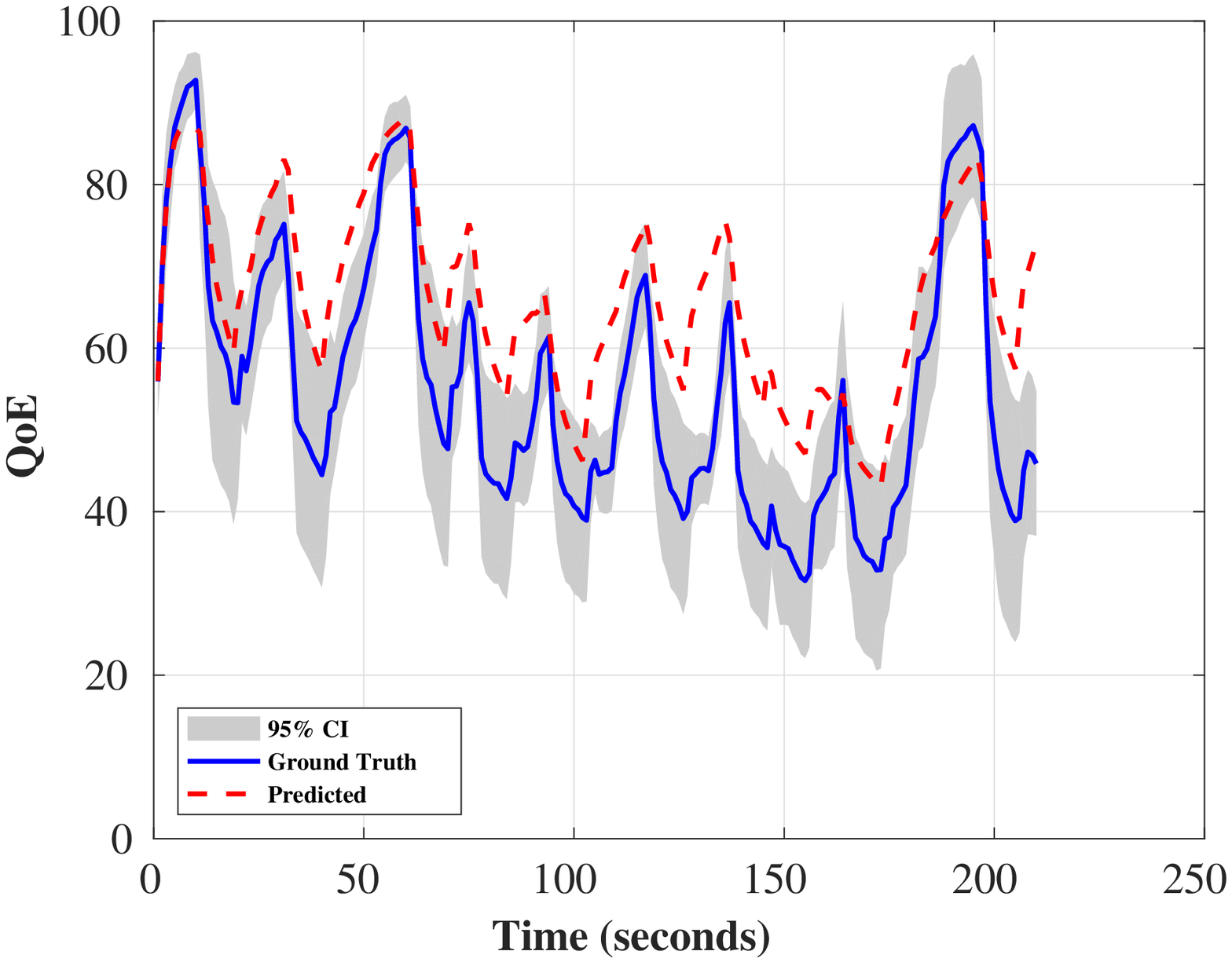}
\hspace{-0.3cm}
\caption{QoE prediction performance of the LSTM-QoE model over the LFOVIA QoE Database \cite{LFOVIA_QoE} with NIQE as the STSQ metric.}
\label{fig:LFOVIA_QoE_LSTM}
\end{figure*}

\begin{figure*}[t]
\centering
\includegraphics[scale=0.33]{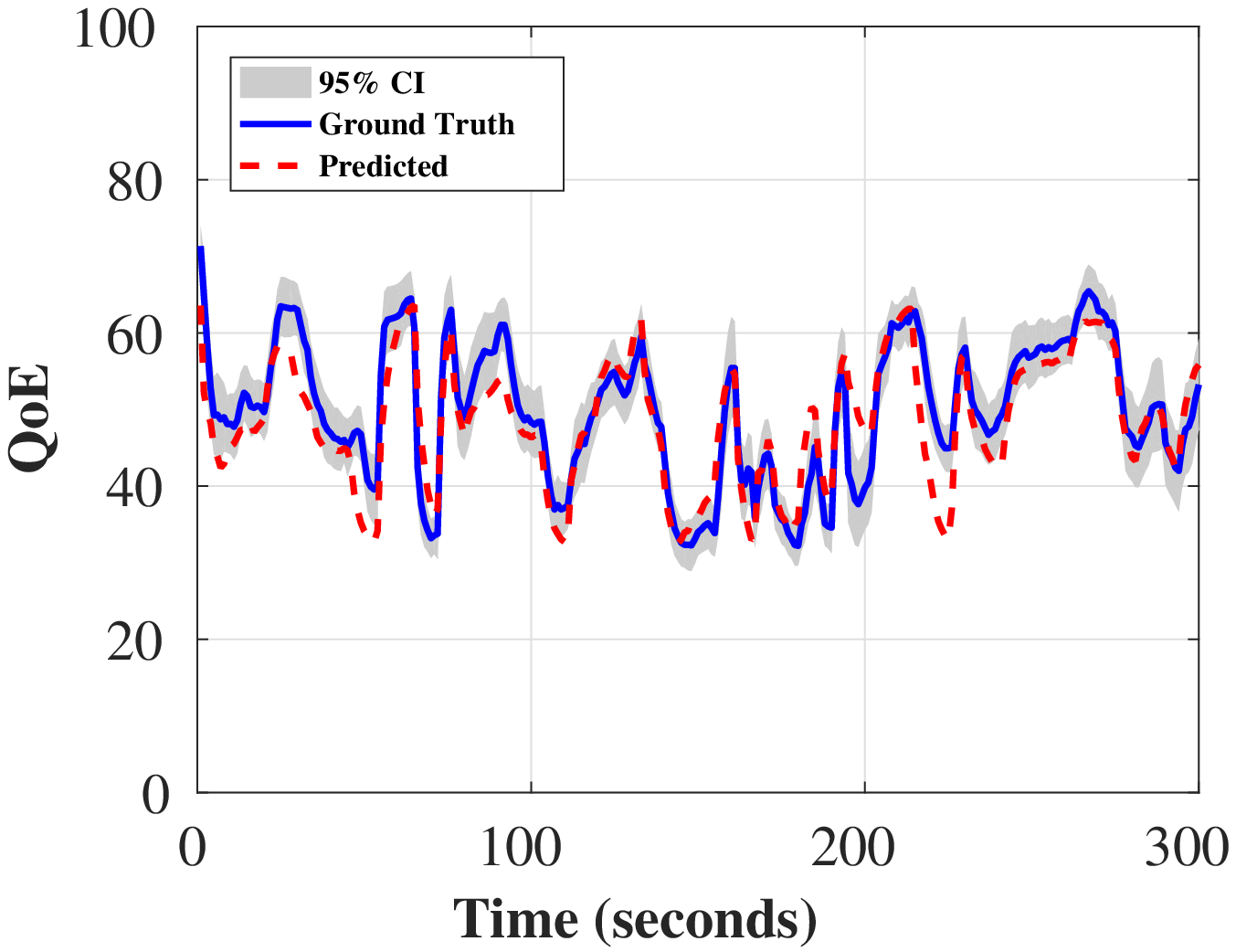}
\hspace{-0.32cm}
\includegraphics[scale=0.33]{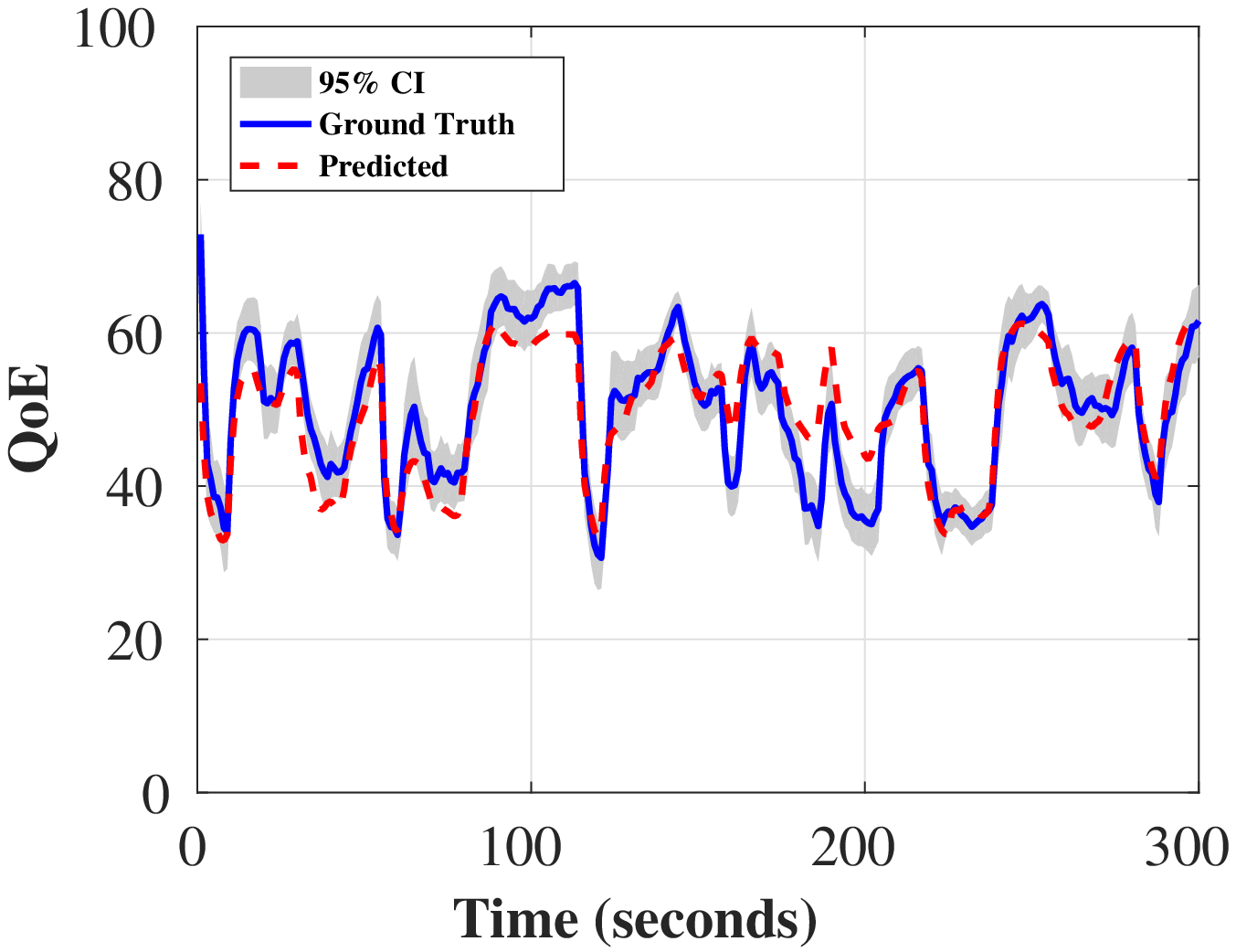}
\hspace{-0.32cm}
\includegraphics[scale=0.33]{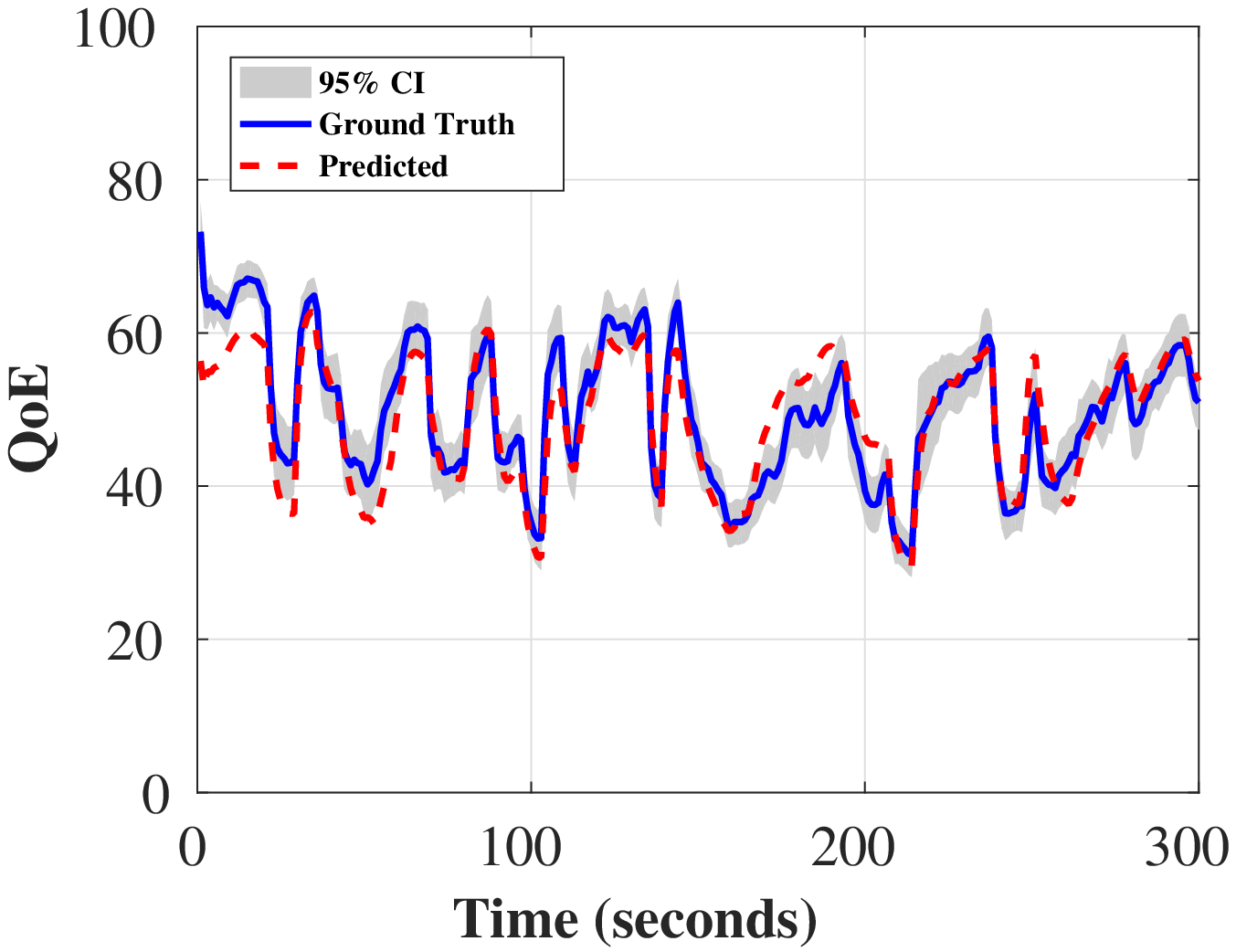}
\hspace{-0.32cm}
\includegraphics[scale=0.33]{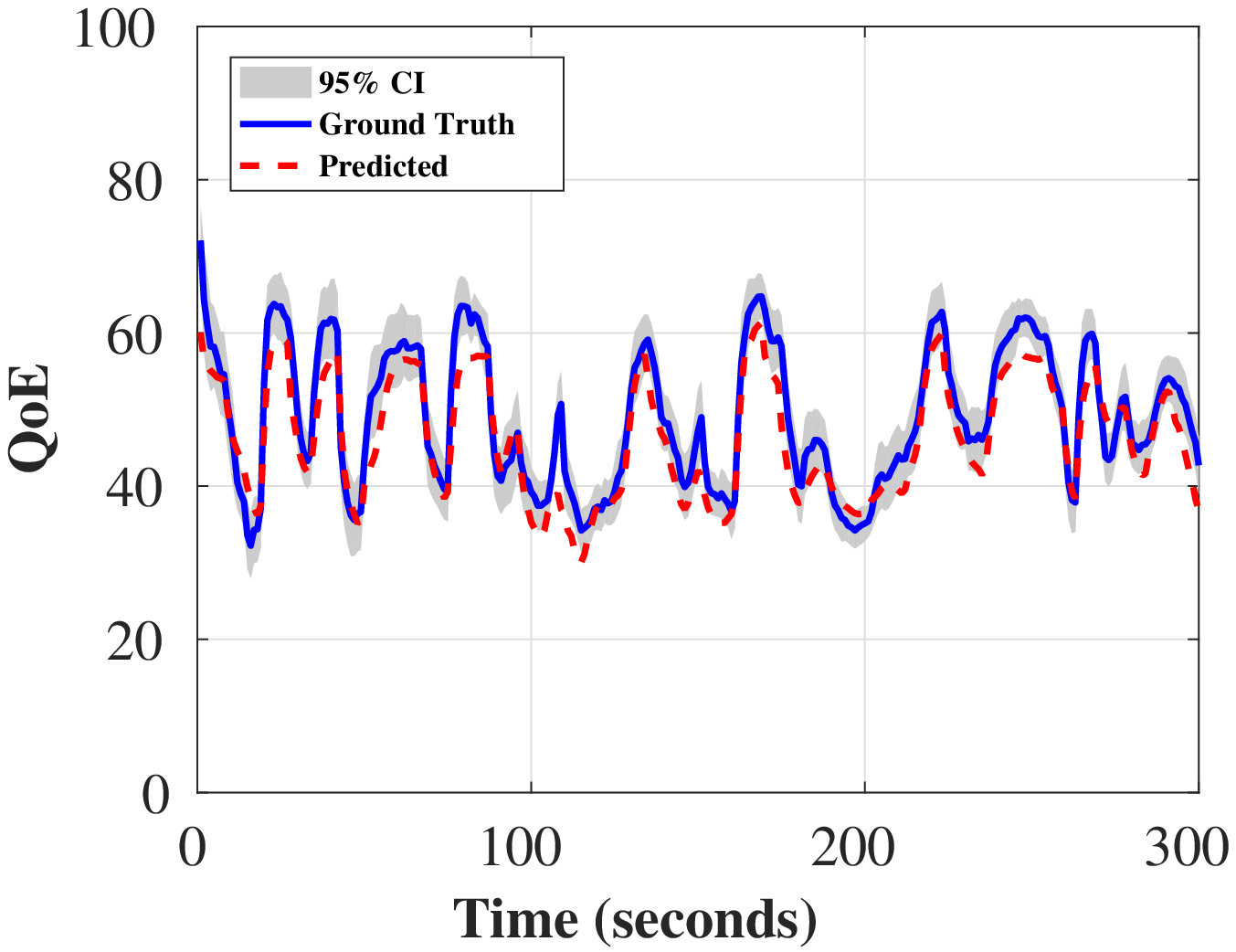}
\hspace{-0.32cm}
\caption{QoE prediction performance of the LSTM-QoE model over the LIVE QoE Database \cite{TVSQ_Chen}  with STRRED as the STSQ metric.}
\label{fig:LIVE_QoE_LSTM}
\end{figure*}

\begin{figure*}[t]
\centering
\includegraphics[scale=0.25]{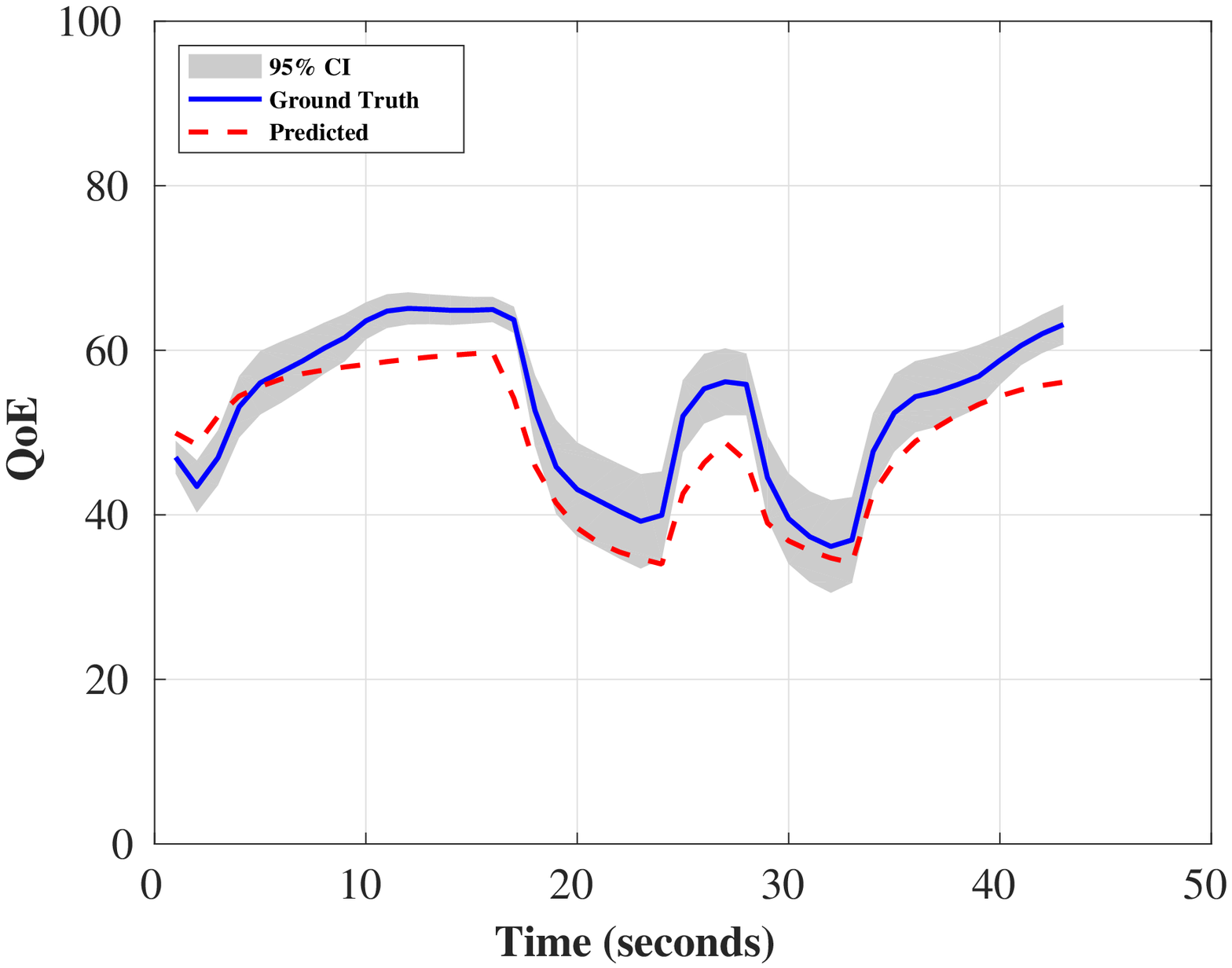}
\hspace{-0.3cm}
\includegraphics[scale=0.25]{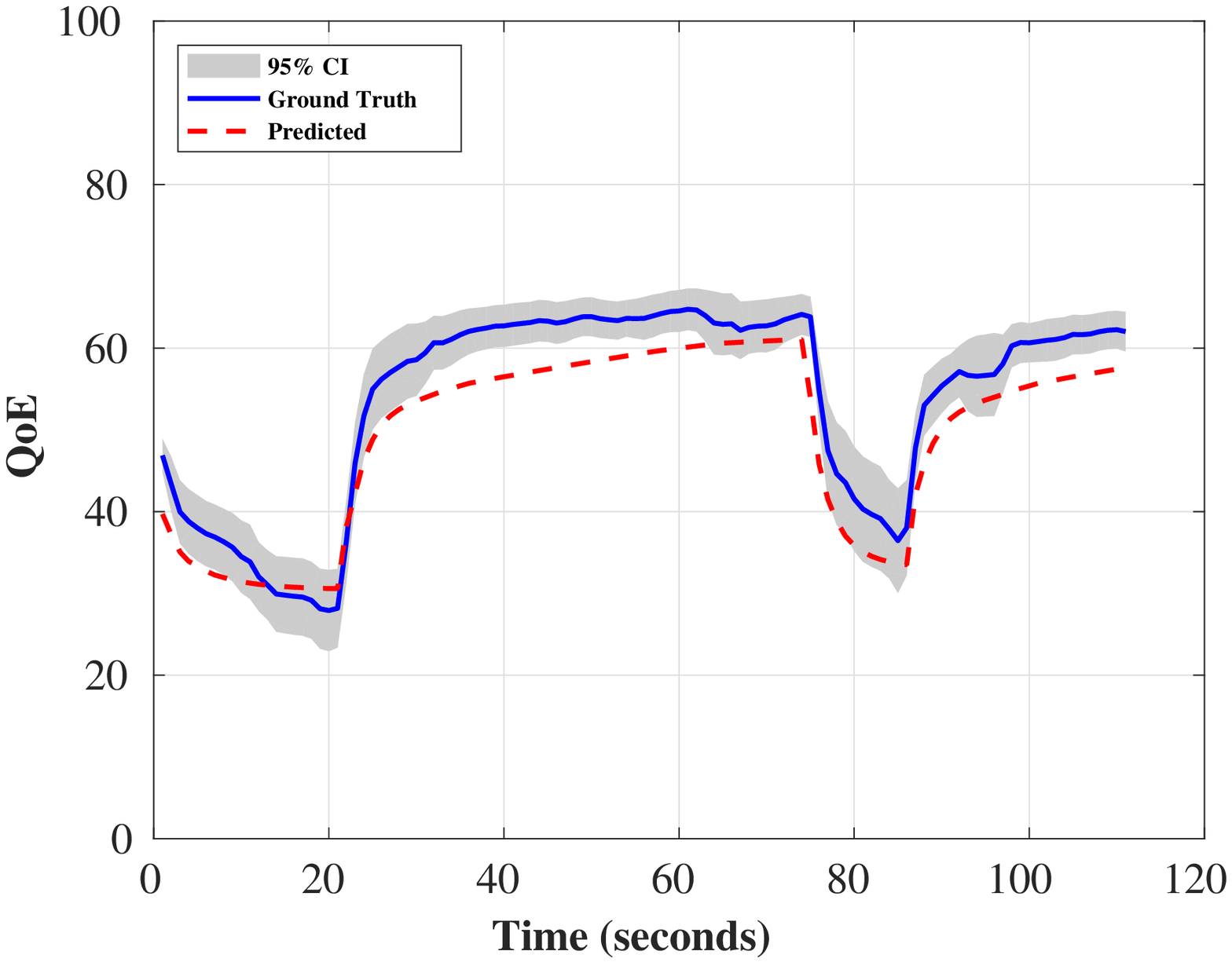}
\hspace{-0.3cm}
\includegraphics[scale=0.25]{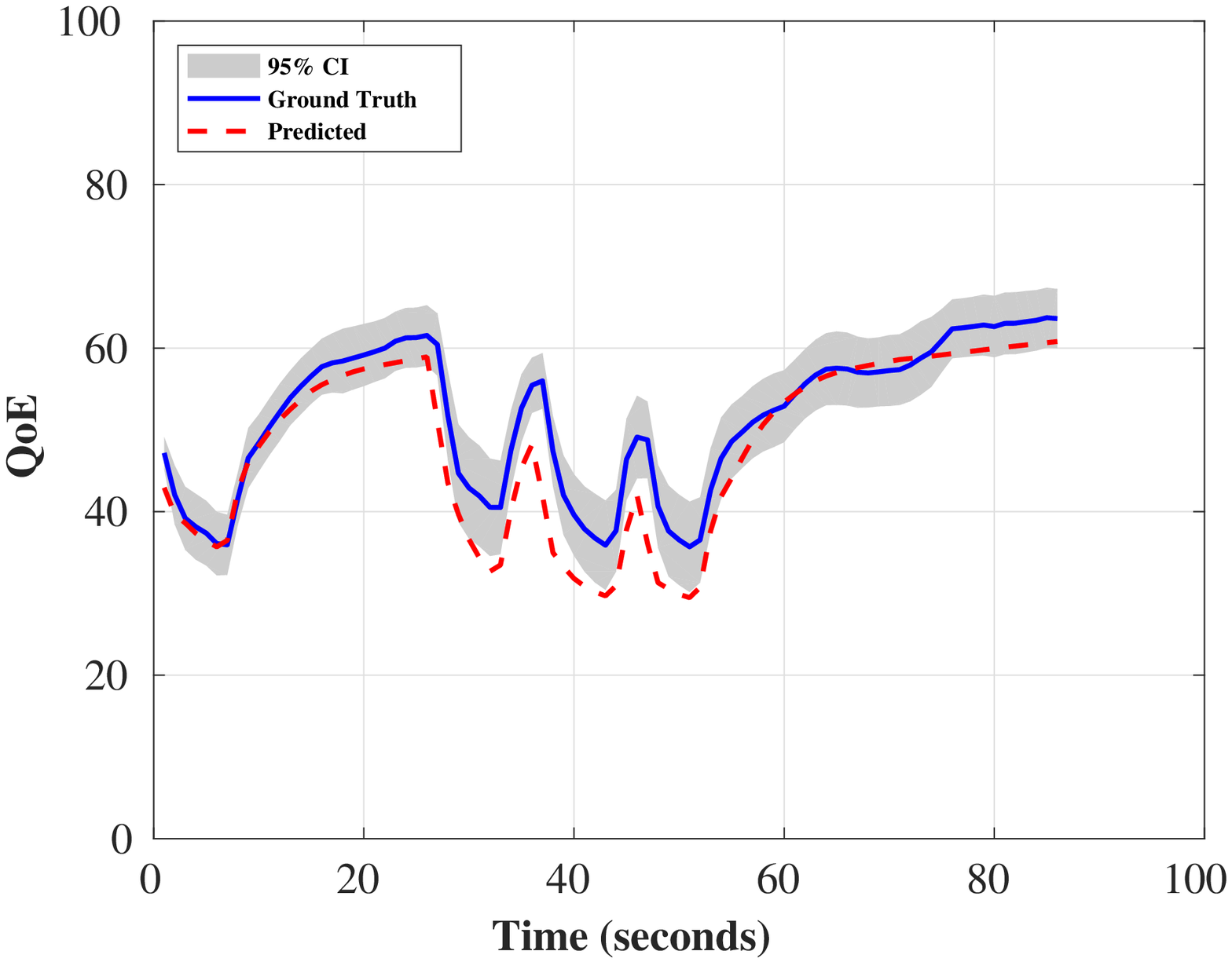}
\hspace{-0.3cm}
\includegraphics[scale=0.25]{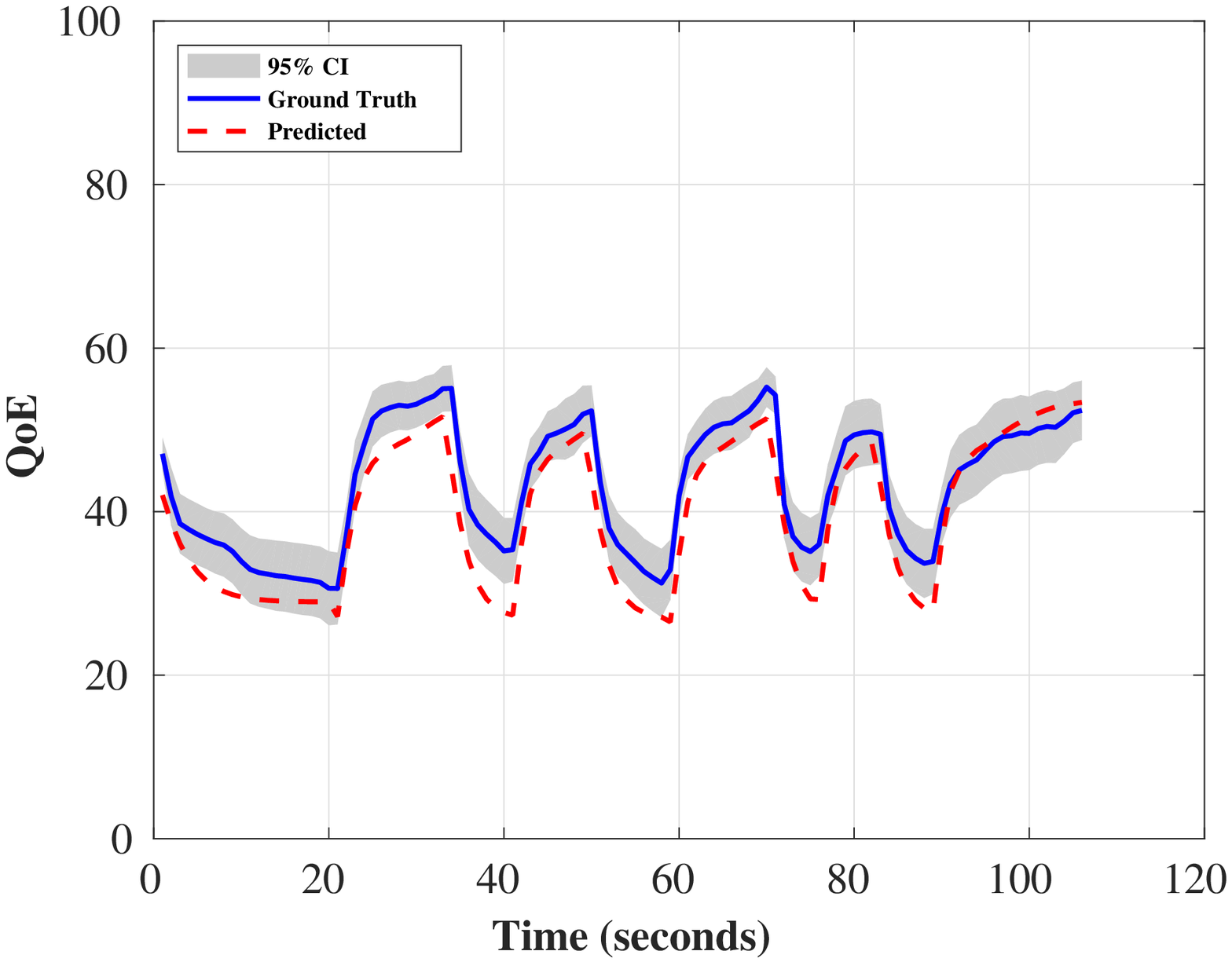}
\hspace{-0.3cm}
\caption{QoE prediction performance of the proposed LSTM-QoE model over the LIVE Mobile Video Stall Database-II \cite{LIVE_Mobile_Stall_II}.}
\label{fig:LIVE_Mobile_Stall_II_LSTM}
\end{figure*}

In this subsection, we investigate the effect of the number of layers and the number of units in the proposed LSTM network for QoE prediction. We vary the number of layers and the number of LSTM units for examining the prediction performance using the features STSQ, PI, and $\textnormal{T}_\textnormal{R}$ on the LIVE Netflix and the LFOVIA QoE Databases.
We consider STRRED \cite{STRRED} for STSQ in this investigation.
We begin with the performance investigation of a single LSTM unit and subsequently increase the number of units and layers in the LSTM network. Fig. \ref{fig:LSTM_layers} illustrates the prediction performance for various configurations of LSTM units and layers. 
We observe that there is a significant improvement in the prediction performance in terms of LCC and OR with the addition of LSTM units and layers to the network. For network configurations involving 2 layers and above, there is a steady increase in the performance upto 10 LSTM units after which the LCC performance begins to saturate. The prediction performance in terms of OR continues to show an improvement beyond 10 units and the improvement begins to saturate beyond 22 LSTM units. 
Although the addition of LSTM layers beyond 2 gives a marginal improvement, we observe that the performance starts diminishing for configurations beyond 5 LSTM layers. 
This could be due to the fact that the network gets deeper with increase in the number of LSTM units and layers and training such a larger network
could be less effective due to potential over-fitting.
Based on a careful examination of LCC and OR performances,
we find that the configuration with 2 LSTM layers and 22 units is the optimal choice of the LSTM network for QoE prediction, i.e., $l$ = 2 and $d$ = 22. Using this configuration, we evaluate the performance of the proposed model on the remaining two databases in the following subsections.
We next discuss the contribution of the individual features for QoE prediction.

\begin{table}[!]
\caption{QoE prediction performance of the LSTM-QoE model over the LIVE Netflix Database \cite{LIVE_Netflix}. The best performing results are indicated in bold.}
\hspace{-0.9cm}
\begin{tabular}{c|c|c|c|c|c} \hline
\label{table:LIVE_Netflix_LSTM}
\textbf{QoE} & \multirow{2}{*}{\textbf{VQA}} & \multirow{2}{*}{\textbf{LCC}} & \multirow{2}{*}{\textbf{SROCC}} & \multirow{2}{*}{\textbf{RMSE}$_\textnormal{n}$\textbf{(}\%\textbf{)}}  & \multirow{2}{*}{\textbf{OR(}\%{\textbf{)}}} \\
\textbf{Model} & & & & \\ \hline

\multirow{3}{*}{LSTM-QoE} & STRRED \cite{STRRED} & \textbf{0.802} & \textbf{0.714} & \textbf{7.78} & 27.39 \\
& MS-SSIM \cite{MSSSIM} & 0.745 & 0.689 & 10.21 & 40.99  \\

& NIQE \cite{NIQE}  & 0.683 & 0.609 & 10.86 & 44.12  \\ \hline

\multirow{3}{*}{NLSS-QoE \cite{NLSS_QoE}} & STRRED \cite{STRRED}  & 0.655 & 0.483 & 16.09 & 69.16 \\
& MS-SSIM \cite{MSSSIM}  & 0.583 & 0.420 & 18.22 & 73.74  \\
& NIQE \cite{NIQE}  & 0.527 & 0.300 & 14.50 & 53.33  \\ \hline

\multirow{3}{*}{NARX \cite{NARX}} & STRRED \cite{STRRED} & 0.621 & 0.557 & 8.52 & \textbf{23.84} \\
& MS-SSIM \cite{MSSSIM} & 0.598 & 0.549 & 10.27 & 25.95 \\
& NIQE \cite{NIQE} & 0.605 & 0.537 & 9.82 & 30.66 \\ \hline

\end{tabular}
\end{table}

\begin{table}[!]
\caption{QoE prediction performance of the LSTM-QoE model over the LFOVIA QoE Database \cite{LFOVIA_QoE}. The best performing results are indicated in bold.}
\hspace{-0.9cm}
\begin{tabular}{c|c|c|c|c|c} \hline
\label{table:LFOVIA_QoE_LSTM}
\textbf{QoE} & \multirow{2}{*}{\textbf{VQA}} & \multirow{2}{*}{\textbf{LCC}} & \multirow{2}{*}{\textbf{SROCC}} & \multirow{2}{*}{\textbf{RMSE}$_\textnormal{n}$\textbf{(}\%\textbf{)}}  & \multirow{2}{*}{\textbf{OR(}\%{\textbf{)}}} \\
\textbf{Model} & & & & \\ \hline

\multirow{3}{*}{LSTM-QoE} & STRRED \cite{STRRED} & 0.800 & 0.730 & 9.56 & 13.72 \\
& MS-SSIM \cite{MSSSIM} & 0.786 & 0.712 & 9.21 & 12.34 \\
& NIQE \cite{NIQE} & \textbf{0.858} & \textbf{0.808} & 8.64 & 11.34 \\ \hline

\multirow{3}{*}{NLSS-QoE \cite{NLSS_QoE}} & STRRED \cite{STRRED} & 0.767 & 0.685 & 7.59 & 8.47 \\
& MS-SSIM \cite{MSSSIM} & 0.781 & 0.680 & 7.37 & 6.78 \\
& NIQE \cite{NIQE} & 0.825 & 0.794 & \textbf{6.97} & \textbf{6.51} \\ \hline

\multirow{3}{*}{SVR-QoE \cite{LFOVIA_QoE}}  & STRRED \cite{STRRED} & 0.686 & 0.648 & 10.44 & 22.87 \\
& MS-SSIM \cite{MSSSIM} & 0.737 & 0.683 & 9.48 & 18.25 \\
& NIQE \cite{NIQE} & 0.797 & 0.750 & 8.32 & 13.64 \\ \hline

\end{tabular}
\end{table}

\begin{table}[!]
\caption{QoE prediction performance of the proposed model over the LIVE QoE Database \cite{TVSQ_Chen}. The best performing results are indicated in bold.}
\hspace{-0.9cm}
\begin{tabular}{c|c|c|c|c|c} \hline
\label{table:LIVE_QoE_LSTM}

\textbf{QoE} & \multirow{2}{*}{\textbf{VQA}} & \multirow{2}{*}{\textbf{LCC}} & \multirow{2}{*}{\textbf{SROCC}} & \multirow{2}{*}{\textbf{RMSE}$_\textnormal{n}$\textbf{(}\%{\textbf{)}}}  & \multirow{2}{*}{\textbf{OR(}\%{\textbf{)}}} \\
\textbf{Model} & & & & \\ \hline

\multirow{3}{*}{LSTM-QoE} & STRRED \cite{STRRED} & \textbf{0.892} & \textbf{0.893} & \textbf{4.55} & \textbf{8.69} \\
& MS-SSIM \cite{MSSSIM} & 0.344 & 0.417 & 10.44 & 42.78 \\

& NIQE \cite{NIQE} & 0.473 & 0.475 & 8.44 & 38.80 \\ \hline

\multirow{3}{*}{NLSS-QoE \cite{NLSS_QoE}} & STRRED \cite{STRRED} & 0.723 & 0.707 & 7.04 & 26.22 \\
& MS-SSIM \cite{MSSSIM} & 0.883 & 0.871 & 4.58 & 11.36 \\
& NIQE \cite{NIQE} & 0.211 & 0.189 & 9.23 & 43.47 \\ \hline

\multirow{3}{*}{HW \cite{TVSQ_Chen}} & STRRED \cite{STRRED} & 0.742 & 0.732 & 7.40 & 32.02 \\
& MS-SSIM \cite{MSSSIM} & 0.727 & 0.705 & 6.70 & 29.11 \\
& NIQE \cite{NIQE} & 0.511 & 0.509 & 8.34 & 36.02 \\ \hline

\end{tabular}
\end{table}

\begin{table}[!]
\caption{QoE prediction performance of the proposed model over the LIVE Mobile Video Stall Database-II \cite{LIVE_Mobile_Stall_II}. The best performing results are indicated in bold.}
\centering
\begin{tabular}{c|c|c|c|c} \hline
\label{table:LIVE_Mobile_Stall_II_LSTM}

\textbf{QoE} & \multirow{2}{*}{\textbf{LCC}} & \multirow{2}{*}{\textbf{SROCC}} & \multirow{2}{*}{\textbf{RMSE}$_\textnormal{n}$\textbf{(}\%\textbf{)}} & \multirow{2}{*}{\textbf{OR(}\%{\textbf{)}}} \\
\textbf{Model} & & & & \\ \hline

LSTM-QoE & \textbf{0.878} & \textbf{0.862} & \textbf{7.08} & \textbf{30.89} \\ \hline

NLSS-QoE \cite{NLSS_QoE} & 0.680 & 0.590 & 9.52 & 42.40 \\ \hline

\end{tabular}
\end{table}

\begin{table}[!]
\caption{QoE prediction performance comparison of LSTM-QoE against TV-QoE \cite{Deepti_contQoE} over the $V_s$ set videos of the LIVE Netflix Database \cite{LIVE_Netflix} with 80/20 split. The best performing results are indicated in bold.}
\centering
\begin{tabular}{c|c|c|c|c} \hline
\label{table:LIVE_Netflix_LSTM_Vs}

\textbf{QoE} & \multirow{2}{*}{\textbf{LCC}} & \multirow{2}{*}{\textbf{SROCC}} & \multirow{2}{*}{\textbf{RMSE}} & \multirow{2}{*}{\textbf{OR(}\%{\textbf{)}}} \\
\textbf{Model} & & & & \\ \hline

LSTM-QoE & \textbf{0.947} & \textbf{0.853} & \textbf{0.238} & \textbf{6.849} \\ \hline

TV-QoE \cite{Deepti_contQoE} & 0.891 & 0.806 & 0.300 & -- \\ \hline

\end{tabular}
\end{table}

\begin{table}[!]
\caption{QoE prediction performance comparison of LSTM-QoE against TV-QoE \cite{Deepti_contQoE} over the $V_c$ set videos of the LIVE Netflix Database \cite{LIVE_Netflix} with 80/20 split. The best performing results are indicated in bold.}
\vspace{0.1in}
\centering
\begin{tabular}{c|c|c|c|c} \hline
\label{table:LIVE_Netflix_LSTM_Vc}

\textbf{QoE} & \multirow{2}{*}{\textbf{LCC}} & \multirow{2}{*}{\textbf{SROCC}} & \multirow{2}{*}{\textbf{RMSE}} & \multirow{2}{*}{\textbf{OR(}\%{\textbf{)}}} \\
\textbf{Model} & & & & \\ \hline

LSTM-QoE & \textbf{0.770} & \textbf{0.787} & \textbf{0.279} & \textbf{18.397} \\ \hline

TV-QoE \cite{Deepti_contQoE} & 0.673 & 0.578 & 0.396 & -- \\ \hline

\end{tabular}
\end{table}

\begin{table}[!]
\caption{QoE prediction performance comparison of LSTM-QoE against TV-QoE \cite{Deepti_contQoE} over the LIVE Mobile Video Stall Database-II \cite{LIVE_Mobile_Stall_II} with 80/20 split. The best performing results are indicated in bold.}
\vspace{0.1in}
\centering
\begin{tabular}{c|c|c|c|c} \hline
\label{table:LIVE_Mobile Stall_II_LSTM_median}

\textbf{QoE} & \multirow{2}{*}{\textbf{LCC}} & \multirow{2}{*}{\textbf{SROCC}} & \multirow{2}{*}{\textbf{RMSE}} & \multirow{2}{*}{\textbf{OR(}\%{\textbf{)}}} \\
\textbf{Model} & & & & \\ \hline

LSTM-QoE & 0.939 & 0.936 & 5.702 & 14.870 \\ \hline

TV-QoE \cite{Deepti_contQoE} & \textbf{0.960} & \textbf{0.944} & \textbf{4.424} & -- \\ \hline

\end{tabular}
\end{table}

\subsection{Feature Contribution}
\label{subsec:feat_comb}

We empirically investigate the contribution of the individual features for QoE prediction. Specifically, we feed the features to the LSTM-QoE network in combination of their subsets and evaluate for their QoE prediction performance on the LIVE Netflix and the LFOVIA QoE Databases. The LSTM network with 2 layers and 22 units is employed with STRRED \cite{STRRED} for STSQ.  Fig. \ref{fig:LSTM_features} illustrates the prediction performance of various feature combinations in terms of LCC and OR. A combination that yields a higher LCC and lower OR is desired. 
It is observed that the best LCC and OR performance is obtained when all the features are employed for QoE prediction. Hence, in our evaluations, we employ all the features, i.e., STSQ, PI, and $\textnormal{T}_\textnormal{R}$ for QoE prediction.
We describe the evaluation of LSTM-QoE and discuss its performance in the following subsection.

\subsection{LSTM-QoE Evaluation}

We train the proposed LSTM-QoE network as described in Section \ref{subsec:QoE_prediction} for evaluation over each database using Keras \cite{Keras}. In all our evaluations, we employ the best LSTM network configuration as determined in Section \ref{subsec:LSTM_layers_units}. 
We employ all the three features, i.e., STSQ, PI and $\textnormal{T}_\textnormal{R}$ for QoE prediction as discussed in Section \ref{subsec:feat_comb}.
We investigate three VQA metrics for STSQ: 1) STRRED \cite{STRRED}, a reduced-reference metric, 2) MS-SSIM \cite{MSSSIM}, a full-reference metric, and 3) NIQE \cite{NIQE}, a no-reference metric.
During training, the data is fed to the network through an input layer with appropriate timesteps as depicted in Fig. \ref{fig:LSTM_architecture}. In the training process, we set timestep = 4 motivated by the III-order temporal dependency employed in NLSS-QoE in \cite{NLSS_QoE}. 
While testing, the QoE $\hat{y}(t)$ is predicted with a granularity of 1 second. Hence, during testing, we perform the prediction every timestep, i.e., timestep~=~1 at the end of the last layer, i.e., the time distributed dense layer.

Figs. \ref{fig:LIVE_Netflix_LSTM}, \ref{fig:LFOVIA_QoE_LSTM}, \ref{fig:LIVE_QoE_LSTM}, and \ref{fig:LIVE_Mobile_Stall_II_LSTM} illustrate the QoE prediction on the considered databases using the proposed LSTM-QoE approach.
The mean QoE prediction performance results for each database are tabulated in the Tables \ref{table:LIVE_Netflix_LSTM}, \ref{table:LFOVIA_QoE_LSTM},  \ref{table:LIVE_QoE_LSTM}, and \ref{table:LIVE_Mobile_Stall_II_LSTM}. 
These Tables also depict the performance of the state-of-the-art QoE models over the respective databases. 
In comparison with the existing models such as NARX on the LIVE Netflix Database \cite{NARX}, SVR-QoE on the LFOVIA QoE Database \cite{LFOVIA_QoE}, and HW on the LIVE QoE Database \cite{TVSQ_Chen}, which are the QoE models proposed on the respective databases, we observe that the proposed LSTM-QoE model provides a superior prediction performance. From Table \ref{table:LIVE_Netflix_LSTM}, we observe that LSTM-QoE outperforms NLSS-QoE and NARX models in terms of LCC, SROCC and $\textnormal{RMSE}_\textnormal{n}$ and yields a competitive performance in terms of OR against NARX. We also observe that STRRED is the best performing VQA metric for measuring STSQ.
From Table \ref{table:LFOVIA_QoE_LSTM}, we observe that NIQE emerges as the best performing VQA metric for STSQ, with LSTM-QoE providing superior performance in terms of LCC and SROCC. However, it should be noted that the performance with STRRED as the STSQ metric is not too inferior as compared to that of NIQE.
From Tables \ref{table:LIVE_QoE_LSTM} and \ref{table:LIVE_Mobile_Stall_II_LSTM}, it can be seen that the LSTM-QoE outperforms the existing QoE models across all the performance measures.

We also compare the median QoE prediction performances obtained by the proposed LSTM-QoE model with that of TV-QoE \cite{Deepti_contQoE} on the LIVE Netflix \cite{LIVE_Netflix} and the LIVE Mobile Video Stall-II \cite{LIVE_Mobile_Stall_II} Databases. 
For a fair comparison, we employ a training-test split of 80/20 as considered in \cite{Deepti_contQoE} for evaluation over both these databases.
Upon the LIVE Netflix Database, we conduct the evaluation on two sets separately, as performed in \cite{Deepti_contQoE}: 1) $V_c$: the set of videos having compression artifacts only and 2) $V_s$: the set of videos having both compression and stalling (rebuffering) artifacts. 
The median QoE prediction performances on $V_s$ and $V_c$ video sets are tabulated in Tables \ref{table:LIVE_Netflix_LSTM_Vs} and \ref{table:LIVE_Netflix_LSTM_Vc}, respectively. A superior QoE prediction performance of LSTM-QoE over TV-QoE \cite{Deepti_contQoE} can be observed from these Tables. Moreover, the prediction performance is consistently superior across LCC, SROCC, and RMSE measures.
The median performance of LSTM-QoE on the LIVE Mobile Video Stall Database-II is tabulated in Table \ref{table:LIVE_Mobile Stall_II_LSTM_median}.
Although the performance of LSTM-QoE on the LIVE Mobile Video Stall Database-II is slightly inferior, it is highly competitive as compared to the TV-QoE \cite{Deepti_contQoE}. 
On the other side, the TV-QoE provides relatively inferior performance over both $V_c$ and $V_s$ sets of the LIVE Netflix Database as compared to that of LSTM-QoE.

For the proposed model, while STRRED as the STSQ measure performs the best on the LIVE Netflix and the LIVE QoE Databases, NIQE emerges as the best performing VQA metric on the LFOVIA QoE Database. 
It must be noted that the LFOVIA QoE Database consists of videos at FHD and UHD resolutions. 
Although the VQA performance of STRRED has been demonstrated over the resolution 768$\times$432 \cite{STRRED}, its performance at higher resolutions such as FHD and UHD is not well studied.
NIQE being a no-reference image quality assessment (IQA) metric is applied frame-by-frame on videos to measure the video quality. The better performance provided by NIQE can be attributed due to its better quality prediction capabilities at higher resolutions.
Hence, we hypothesize that the difference in the QoE performance across different STSQ measures is due to the dependency of the STSQ metrics on the video resolution. 
This indicates that there is a scope for more efficient VQA algorithms that work consistently well across resolutions.
Nevertheless, from the Tables \ref{table:LIVE_Netflix_LSTM}, \ref{table:LFOVIA_QoE_LSTM}, and \ref{table:LIVE_QoE_LSTM}, it can be inferred that STRRED can serve as a good metric for STSQ measurement.
Furthermore, the proposed model provides the flexibility to choose the appropriate VQA for QoE prediction. The results also demonstrate the effectiveness of the chosen features for QoE prediction.
Further, we would like to highlight that the best performing STSQ metrics observed in the proposed model concur with those of the QoE models reported over the respective databases \cite{NARX,LFOVIA_QoE,TVSQ_Chen}.

We would also like to note that while testing, the QoE computation using the proposed LSTM network is performed in a feedforward fashion, similar to as performed in the state-of-the-art QoE models such as NARX \cite{NARX} and SVR-QoE \cite{LFOVIA_QoE}. The QoE computational complexity in these models is determined by the computational complexity of STSQ, which is in turn determined by the VQA method employed for computing STSQ. However, for applications such as on-demand video streaming, the STSQs can be computed offline and can be readily made available in order to facilitate QoE computation in real time.

In summary, the QoE prediction performance offered by the proposed model is superior and consistent across databases. 
The results illustrate that the proposed LSTM-QoE network is capable of capturing the complex temporal dependencies involved in the QoE process, 
thereby demonstrating its efficacy in modeling the non-Markovian QoE dynamics.
Thus, we infer that the LSTM-QoE is a highly efficient and an effective model for QoE prediction.

We next review the proposed approach from the perspective of state space and discuss the connection between them.

\subsection{LSTM-QoE: A State Space Perspective}
\label{sec:SSL}

We have seen the excellent performance offered by the proposed LSTM model for QoE prediction. Such a performance is attributed to the capability of LSTMs in modeling the non-Markovian QoE dynamics by capturing the long and short-term dependencies through the dynamically evolving internal states.
In \cite{SSL}, it is shown that the LSTMs can be modeled as a state space (referred as State Space LSTM (SSL)) using Sequential Monte Carlo inference. The SSLs provide a state space interpretation for modeling the nonlinear non-Markovian dynamics of LSTMs.
On the other hand, the NLSS-QoE proposed in \cite{NLSS_QoE} is a nonlinear QoE prediction model based on the conventional state space approach. It is shown in \cite{NLSS_QoE} that the NLSS-QoE model provides superior QoE prediction over the existing models on the LIVE Netflix \cite{LIVE_Netflix} and the LFOVIA QoE \cite{LFOVIA_QoE} Databases. Such a superior performance is attributed to the model design wherein, the non-Markovian QoE dynamics are captured explicitly through the states.
Further, the state space in NLSS-QoE is constituted by subsets of states that are distinctly controlled by each of the previous `$r$' input features. The number of states in each subset is determined by an empirically chosen order $r$ that explicitly models the non-Markovian dynamics.
However, such a fixed choice of the model order could be stringent and may not effectively capture the dependencies and the hysteresis effects involved in the QoE process. 
This drawback of enforced temporal dependency in the NLSS-QoE model has been overcome in the proposed LTSM-QoE model where the LSTM network learns these dependencies during the training process.
Although there are a fixed number of states in LSTM-QoE as determined by the size of the LSTM network, i.e., the parameters $l$ and $d$, the LSTM latent states $\textbf{c}(t)$ capture the QoE dynamics implicitly through the state transitions as described in (\ref{eq:LSTM_2}). 
Further, we would like to note that the LSTM-QoE does not need explicit state initialization for QoE prediction, whereas, appropriate state initialization is crucial for the NLSS-QoE model.

The efficacy of LSTM-QoE over NLSS-QoE for QoE prediction can also be attributed to the cascaded LSTM nonlinearities in the LSTM network. In NLSS-QoE, a single input nonlinearity drives the linear state space. Whereas, in LSTM-QoE, the nonlinearities are imposed at multiple stages in each LSTM unit 
before feeding the next unit. Such a capability of modeling complex nonlinearities as well as the memory effects through latent states makes the proposed LSTM-QoE highly efficient for QoE prediction.
This is evident from the LSTM-QoE model's performances illustrated in Tables \ref{table:LIVE_Netflix_LSTM}, \ref{table:LFOVIA_QoE_LSTM}, \ref{table:LIVE_QoE_LSTM} and \ref{table:LIVE_Mobile_Stall_II_LSTM} as compared to that of the NLSS-QoE.
Nevertheless, based on the effectiveness and the superior performance offered by the NLSS-QoE and LSTM-QoE models over the existing models, we infer that the state space approaches have immense potential and offer a promising direction for performing effective continuous QoE prediction.

The prediction of overall QoE using the continuous QoE scores is discussed in the following subsection.

\subsection{Overall QoE Performance}
\label{sec:overall_perf}

In this subsection, we investigate whether the overall QoE of the user can be predicted using the continuous QoE scores.
In addition to the continuous QoE scores, the continuous QoE databases also provide the overall QoE obtained at the end of each video during the subjective study.
We use these scores as the ground truth for validating the predicted overall QoE scores. 
In this investigation, we consider the ground truth continuous QoE scores and pool them to predict the overall QoE.
We explore two strategies for pooling the continuous QoE scores for overall QoE prediction. They are 1) mean pooling and 2) median pooling. The pooled continuous QoE scores are correlated against the ground truth overall QoE scores for prediction performance evaluation.
The performance results are tabulated in the Table \ref{table:overall_QoE_prediction}.
It is observed that the overall QoE prediction performance is good under both the pooling strategies, with mean pooling performing slightly better than the median pooling. 
Therefore, we infer that the mean/median pooling strategies on the continuous QoE scores can be effectively used to predict the overall QoE of the user. 
It is interesting to note that the efficacy of the overall QoE prediction depends on the effectiveness of the continuous QoE prediction. Thus, the continuous QoE prediction is useful in providing insights for understanding the overall experience of the user at the end of a video session.

\begin{table}[t]
\caption{Predicting the overall QoE from the continuous QoE scores using different pooling strategies.}
\centering

\begin{tabular}{c|c|c|c} \hline
\label{table:overall_QoE_prediction}

\textbf{Database} & \textbf{Pooling} & \textbf{LCC} & \textbf{SROCC} \\ \hline

\multirow{2}{*}{LIVE Netflix \cite{LIVE_Netflix}} & mean & 0.985 & 0.971  \\ 
& median & 0.966 & 0.933  \\ \hline

\multirow{2}{*}{LFOVIA QoE \cite{LFOVIA_QoE}} & mean & 0.957 & 0.900  \\ 
& median & 0.907  & 0.875  \\ \hline

\multirow{2}{*}{LIVE Mobile Video Stall-II \cite{LIVE_Mobile_Stall_II}} & mean & 0.931 & 0.915  \\ 
& median & 0.927 & 0.921  \\ \hline

\end{tabular}
  
\end{table}

\section{Conclusions}
\label{sec:conclusions}

In this paper, we proposed LSTM-QoE, a novel method for continuous video QoE evaluation. 
The proposed model consists of an LSTM network for capturing the complex temporal dependencies involved in the non-Markovian dynamics of the QoE process. 
The QoE prediction using the proposed model was performed using a set of carefully selected QoE determining features.
A comprehensive evaluation of the proposed model was conducted on four publicly available continuous QoE databases and it was shown that the LSTM-QoE provides an excellent prediction performance across all the databases. It was also observed that the LSTM-QoE outperforms the state-of-the-art QoE prediction models.
Based on the performance of LSTM-QoE and NLSS-QoE, we found that the state space approaches are effective for QoE modeling and possess an immense potential for efficient QoE prediction.
Finally, an overall QoE prediction performance analysis showed that the mean and the median continuous QoE pooling strategies are effective for quantifying the overall QoE of the users.
In future, we intend to develop a highly robust universal QoE predictor that can provide an excellent prediction performance on existing and upcoming QoE databases and across diverse scenarios of video streaming.

\bibliographystyle{IEEEtran}
\bibliography{LSTM_QoE_arxiv}
\end{document}